\documentclass[twocolumn,preprintnumbers,nofootinbib,amsmath,amssymb,floatfix]{revtex4}

\usepackage{graphicx}
\usepackage{dcolumn}
\usepackage{bm}
\usepackage{ifpdf}
\usepackage[tight]{subfigure}

\newif\ifcomment
\newif\ifprint
\printtrue

\ifpdf
\ifprint
 \usepackage[linkbordercolor={0 0 .8},%
             citebordercolor={.8 0 0},%
             urlbordercolor={.8 0 .8},%
             raiselinks=true,%
             pdfborder={0 0 1 [3]}]{hyperref}
\else
 \usepackage[pdftex,backref,pagebackref,colorlinks,
             linkcolor=blue,citecolor=blue,anchorcolor=blue,
             pagecolor=blue,urlcolor=red,a4paper]{hyperref}
\fi
\pdfoutput=1        
\pdfcatalog{/PageMode(/UseOutlines)} 

\else 

\ifprint
 \usepackage[linkbordercolor={0 0 .8},%
             citebordercolor={.8 0 0},%
             urlbordercolor={.8 0 .8}]
             {hyperref}
\else
 \usepackage[dvips,backref,pagebackref,colorlinks,
             linkcolor=blue,citecolor=blue,anchorcolor=blue,
             pagecolor=blue,urlcolor=red,a4paper]{hyperref}
\fi
\hypersetup{
  pdfauthor   = {},
  pdftitle    = {},
  pdfcreator  = {},
  pdfproducer = {},
  pdfsubject  = {},
  pdfkeywords = {}
 }
\fi

\def\prap         {pseudorapidity}
\def\Prap         {Pseudorapidity}

\newcommand {\deta}		{\ensuremath{\Delta\eta}}
\newcommand {\dphi}		{\ensuremath{\Delta\phi}}

\newcommand {\V}[1]{{\mathbf V}_{#1\Delta}}

\newcommand {\mean}[1]{\left\langle #1 \right\rangle}

\newcommand {\der}{{\rm{d}}}

\def\etal         {et al.}

\newcommand {\snn}      {\sqrt{s_{\scriptscriptstyle{{\rm NN}}}}}

\newcommand {\snntwo}   {\mbox{\ensuremath{\snn  =}\ 200\ GeV}}

\newcommand {\abs}[1]   {\ensuremath{\left| #1 \right|}}

\newcommand {\fig}[1]{Fig.~\ref{#1}}

\newcommand {\Fig}[1]{Figure~\ref{#1}}

\newcommand {\eq}[1]{Eq.~\ref{#1}}
\newcommand {\eqs}[2]{Eqs.~\ref{#1} and~\ref{#2}}

\newcommand {\pt}       {\ensuremath{p_{\mathrm{T}}}}

\newcommand {\Npart}    {\ensuremath{N_{\rm part}}}

\newcommand {\gev}   {\mbox{${\rm GeV}$}}

\newcommand {\mom}   {\mbox{\rm GeV$\kern-0.15em /\kern-0.12em c$}}
\newcommand {\mmom}  {\mbox{\rm MeV$\kern-0.15em /\kern-0.12em c$}}
\newcommand {\mass}  {\mbox{\rm GeV$\kern-0.15em /\kern-0.12em c^2$}}
\newcommand {\mmass} {\mbox{\rm MeV$\kern-0.15em /\kern-0.12em c^2$}}

\newcommand {\pp}    {\mbox{p+p}}

\newcommand {\tria}{\varepsilon_{3}}
\newcommand {\ecc}{\varepsilon_{2}}

\DeclareMathOperator{\atantwo}{atan2}

\begin{document}

\title{Collision geometry fluctuations and triangular flow in heavy-ion collisions}

\author{
B.Alver, G.Roland\\
\vspace{3mm}
\small
 Laboratory for Nuclear Science, Massachusetts Institute of Technology,
 Cambridge, MA 02139-4307, USA\\
}

\begin{abstract}\noindent
  We introduce the concepts of participant triangularity and
  triangular flow in heavy-ion collisions, analogous to the
  definitions of participant eccentricity and elliptic flow.  The
  participant triangularity characterizes the triangular anisotropy of
  the initial nuclear overlap geometry and arises from event-by-event
  fluctuations in the participant-nucleon collision points. In studies
  using a multi-phase transport model (AMPT), a triangular flow signal
  is observed that is proportional to the participant triangularity
  and corresponds to a large third Fourier coefficient in two-particle
  azimuthal correlation functions.  Using two-particle azimuthal
  correlations at large \prap\ separations measured by the PHOBOS and
  STAR experiments, we show that this Fourier component is also
  present in data.  Ratios of the second and third Fourier
  coefficients in data exhibit similar trends as a function of
  centrality and transverse momentum as in AMPT calculations.  These
  findings suggest a significant contribution of triangular flow to
  the ridge and broad away-side features observed in data.  Triangular
  flow provides a new handle on the initial collision geometry and
  collective expansion dynamics in heavy-ion collisions.

  \vspace{3mm}
\noindent 
\end{abstract}
\maketitle

\section{Introduction}
Studies of two-particle azimuthal correlations have become a key tool
in characterizing the evolution of the strongly interacting medium
formed in ultra-relativistic nucleus-nucleus collisions.
Traditionally, the observed two-particle azimuthal correlation
structures are thought to arise from two distinct
contributions. The dominant one is the ``elliptic flow'' term, related
to anisotropic hydrodynamic expansion of the medium from an
anisotropic initial state \cite{Ackermann:2000tr, Adcox:2002ms,
  Adler:2003kt, Adams:2003am, Adams:2004bi, Back:2004zg, Back:2004mh,
  Alver:2006wh, Adare:2006ti}.  In addition, one observes so-called
``non-flow'' contributions from, e.g., resonances and jets, which may
be modified by their interactions with the medium~\cite{Adler:2002tq, Adams:2006yt, :2008cqb}.  

The strength 
of anisotropic flow is usually quantified with a Fourier decomposition of
the azimuthal distribution of observed particles relative to the
reaction plane~\cite{Voloshin:1994mz}. The experimental observable
related to elliptic flow is the second Fourier coefficient, ``$v_2$.''
The elliptic flow signal has been studied extensively in Au+Au
collisions at RHIC as a function of \prap, centrality, transverse
momentum, particle species and center of mass energy~\cite{
  Adler:2003kt, Adams:2003am, Adams:2004bi, Back:2004zg, Back:2004mh}.
The centrality and transverse momentum dependence of $v_2$ has been
found to be well described by hydrodynamic calculations, which for a
given equation of state, can be used to relate a given initial energy
density distribution to final momentum distribution of produced
particles~\cite{Kolb:2000fha}. In these calculations, the $v_2$ signal
is found to be proportional to the eccentricity, $\ecc$, of the
initial collision region defined by the overlap of the colliding
nuclei~\cite{Ollitrault:1992bk}.
Detailed comparisons of the observed elliptic flow effects 
with hydrodynamic calculations have led to the conclusion that a new state
of strongly interacting matter with very low shear viscosity,
compared to its entropy density, has been created in these collisions~\cite{Kolb:2000fha,
  Adcox:2004mh, Back:2004je, Adams:2005dq}. 

Measurements of non-flow correlations in heavy-ion collisions, in
comparison to corresponding studies in \pp\ collisions, provide
information on particle production mechanisms~\cite{Alver:2008gk} and
parton-medium interactions~\cite{Adler:2002tq, Adams:2006yt,
  :2008cqb}.  Different methods have been developed to account for the
contribution of elliptic flow to two-particle correlations in these
studies of the underlying non-flow correlations~\cite{Adler:2002tq,
  Adams:2004pa, Ajitanand:2005jj, Alver:2008gk, Trainor:2007fu}. The most commonly
used approach is the zero yield at minimum method (ZYAM), where one
assumes that the associated particle yield correlated with the trigger
particle is zero at the minimum as a function of $\dphi$ after
elliptic flow contribution is taken out~\cite{Ajitanand:2005jj}. The
ZYAM approach has yielded rich correlation structures at $\dphi
\approx 0^{\circ}$ and $\dphi\approx120^{\circ}$ for different $\pt$
ranges~\cite{Adams:2005ph, Adare:2007vu, Alver:2009id, Abelev:2009qa}.
These structures, which are not observed in \pp\ collisions at the
same collision energy, have been referred to as the ``ridge'' and
``broad away-side'' or ``shoulder''.  
The same correlation structures have been found to be present in Pb+Au
collisions at $\snn=17.4~\gev$ at the SPS~\cite{Adamova:2009ah}.
Measurements at RHIC have shown that these structures extend out to
large \prap\ separations of $\deta>2$, similar to elliptic flow
correlations~\cite{Alver:2009id}. The ridge and broad
away-side structures have been extensively studied
experimentally~\cite{Adams:2005ph, Adare:2007vu, Alver:2009id,
  Abelev:2009qa, :2008cqb, :2008nda} and various theoretical models
have been proposed to understand their origin~\cite{Wong:2008yh,
  Pantuev:2007sh, Gavin:2008ev, Dumitru:2008wn, Ruppert:2007mm,
  Pruneau:2007ua, Hwa:2009bh, Takahashi:2009na}.  A recent review of
the theoretical and experimental results can be found
in~\cite{Nagle:2009wr}.

\begin{figure*}[t]
  \centering
  \subfigure{
    \includegraphics[width=0.31\textwidth]{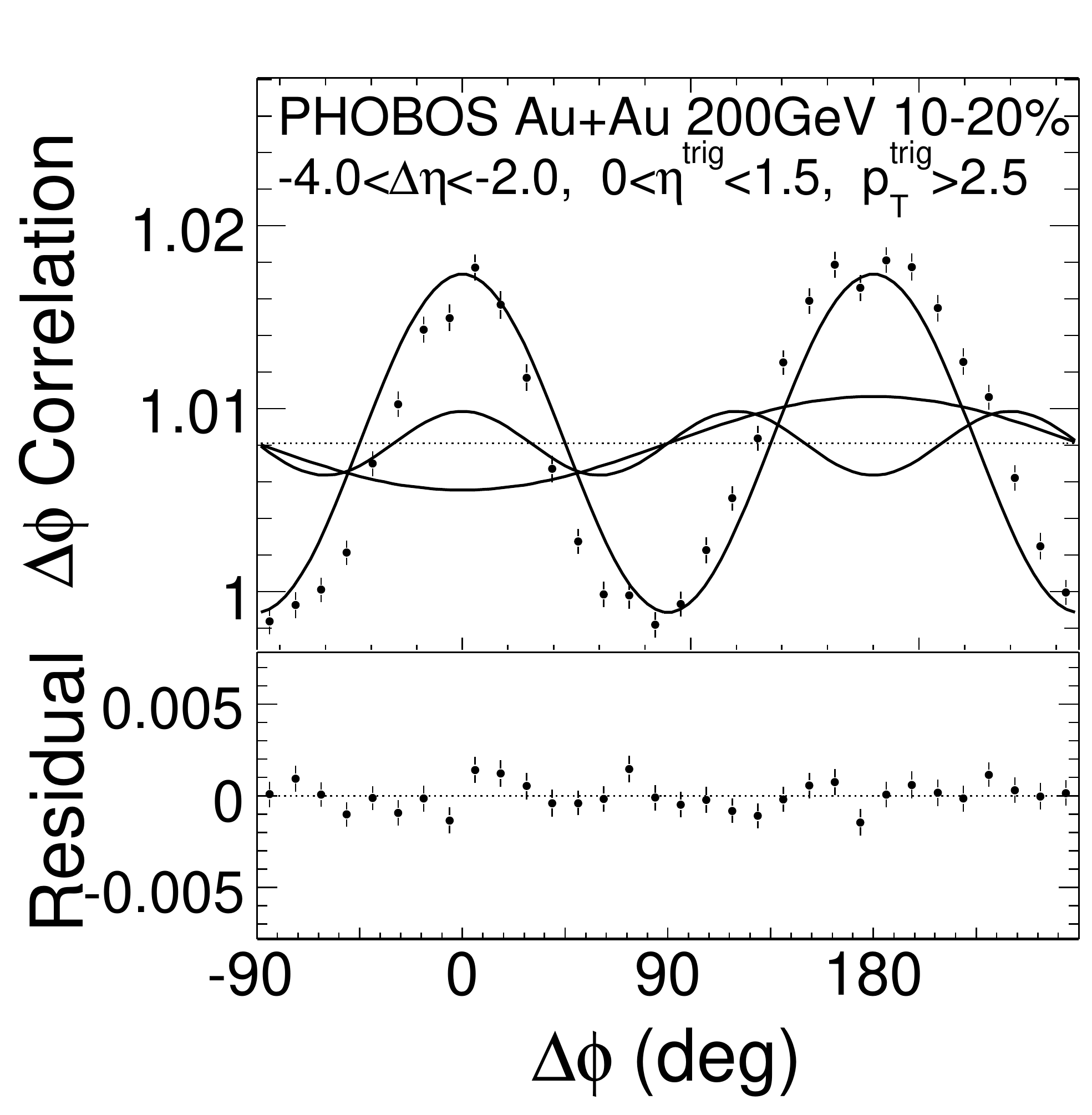}
    \label{fig:corrs1Dsamepad_Ed_cen13}
  }
  \subfigure{
    \includegraphics[width=0.31\textwidth]{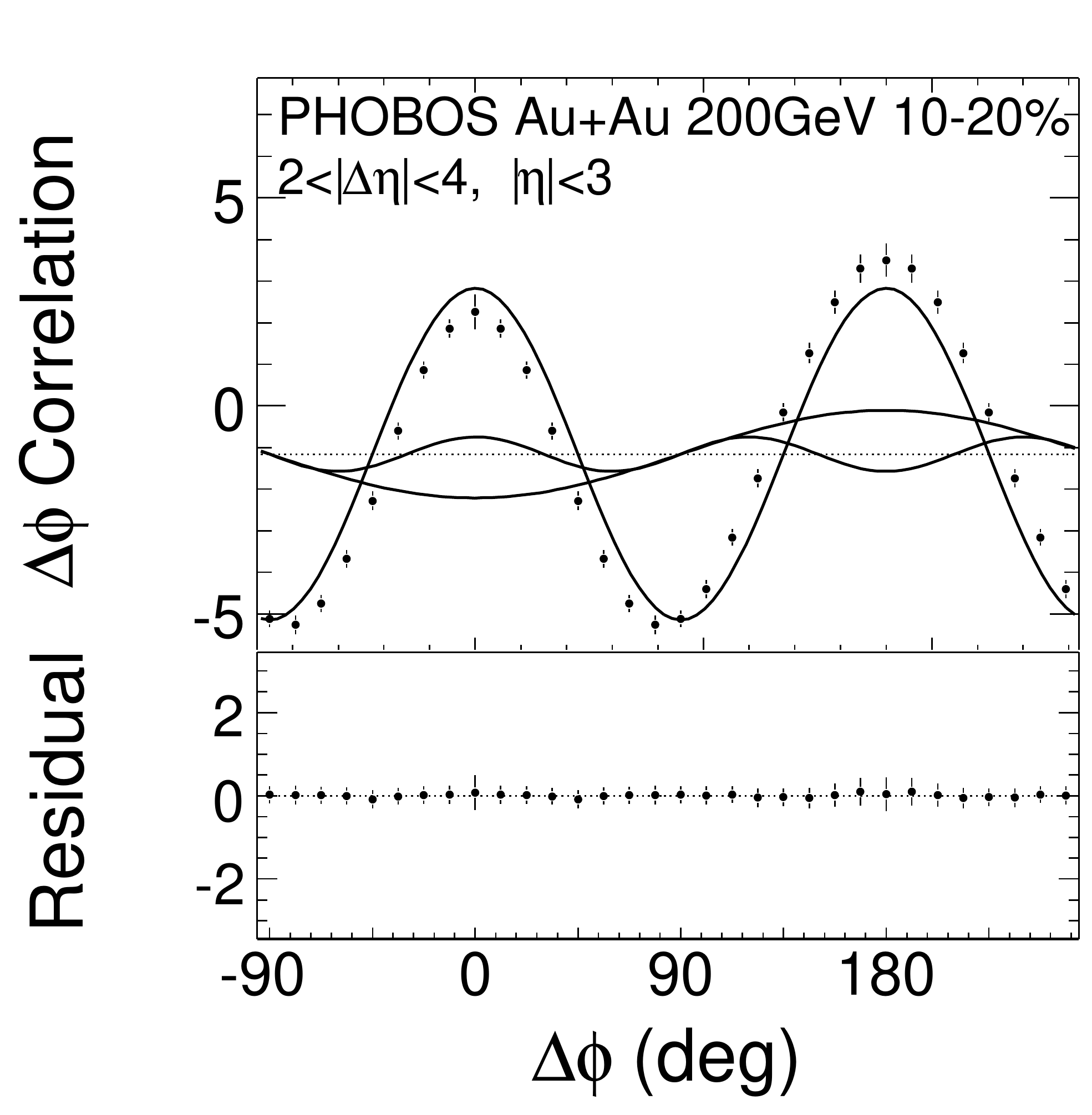}
    \label{fig:corrs1Dsamepad_Wei_cen13}
  }
  \subfigure{
    \includegraphics[width=0.31\textwidth]{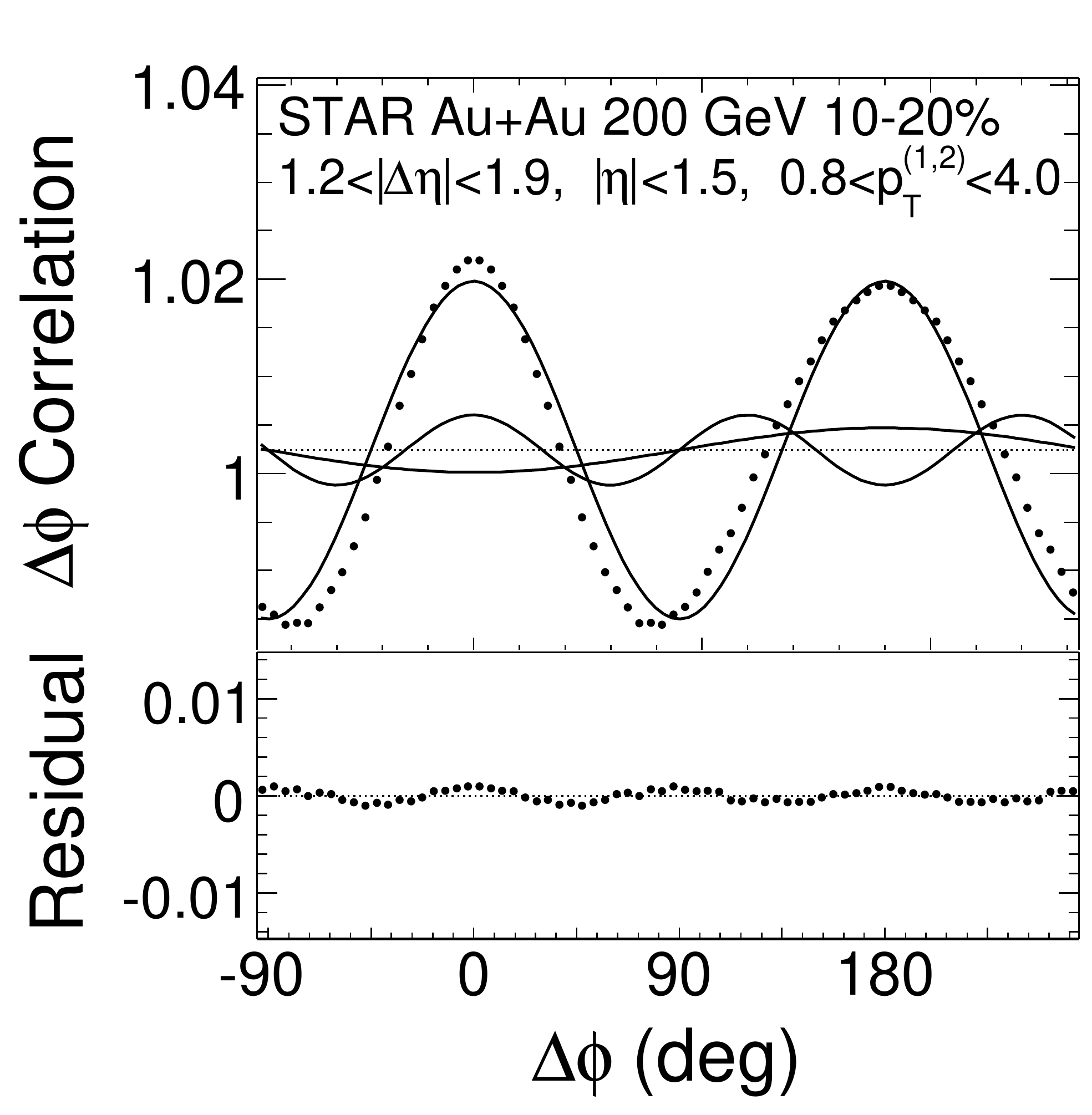}
    \label{fig:corrs1Dsamepad_Trainor_cen1}
  }
  \caption{Top: azimuthal correlation functions for mid-central
    (10-20\%) Au+Au collisions at \snntwo\ obtained from projections
    of two-dimensional $\deta,\dphi$ correlation measurements by
    PHOBOS~\cite{Alver:2009id, Alver:2008gk} and
    STAR~\cite{Abelev:2008un}. The transverse momentum and
    \prap\  ranges are indicated on the figures. Errors bars
    are combined systematic and statistical errors.  The first three
    Fourier components are shown in solid lines. Bottom: the residual
    correlation functions after the first three Fourier components are
    subtracted.}
  \label{fig:corrs1d} 
\end{figure*}

In this paper, we propose that the observed ridge and broad away-side
features in two-particle correlations may be due to an average
triangular anisotropy in the initial collision geometry which is
caused by event-by-event fluctuations and which leads to a triangular
anisotropy in azimuthal particle production through the collective
expansion of the medium. It was shown that, in the NEXSPHERIO
hydrodynamic model, ridge and broad away-side structures in two
particle correlations arise if non-smooth initial conditions are
introduced~\cite{Takahashi:2009na}. Sorensen has suggested that
fluctuations of the initial collision geometry may lead to higher
order Fourier components in the azimuthal correlation function through
collective effects~\cite{Sorensen}.  An analysis of higher order
components in the Fourier decomposition of azimuthal particle
distributions, including the odd terms, was proposed by
Mishra~\etal\ to probe superhorizon fluctuations in the thermalization
stage~\cite{Mishra:2007tw}.  In this work, we show that the second and
third Fourier components of two-particle correlations may be best
studied by treating the components of corresponding initial geometry
fluctuations on equal footing. To
reduce contributions of non-flow correlations, which are most
prominent in short \prap\ separations, we focus on azimuthal
correlations at long ranges in \prap. We show that the ridge and broad
away-side structures can be well described by the first three
coefficients of a Fourier expansion of the azimuthal correlation
function
\begin{equation}
  \frac{\der N^{\text{pairs}}}{\der \dphi} = \frac{N^{\text{pairs}}}{2\pi} \left(1+\sum_{n} 2\V{n}\cos(n\dphi)\right),
\end{equation}
where the first component, $\V{1}$\footnote{Note the distinction between
  $\V{n}$ and $v_n$. See \eqs{eq:vnflow1}{eq:vnflow} for details.}, is
understood to be due to momentum conservation and directed flow and
the second component $\V{2}$ is dominated by the contribution from
elliptic flow.  Studies in a multi-phase transport model
(AMPT)~\cite{Lin:2004en} suggest that not only the elliptic flow term,
$\V{2}$, but also a large part of the correlations measured by the
$\V{3}$ term, arises from the hydrodynamic expansion of the medium.

\begin{figure*}[t]
  \centering
  \subfigure{
    \includegraphics[width=0.45\textwidth]{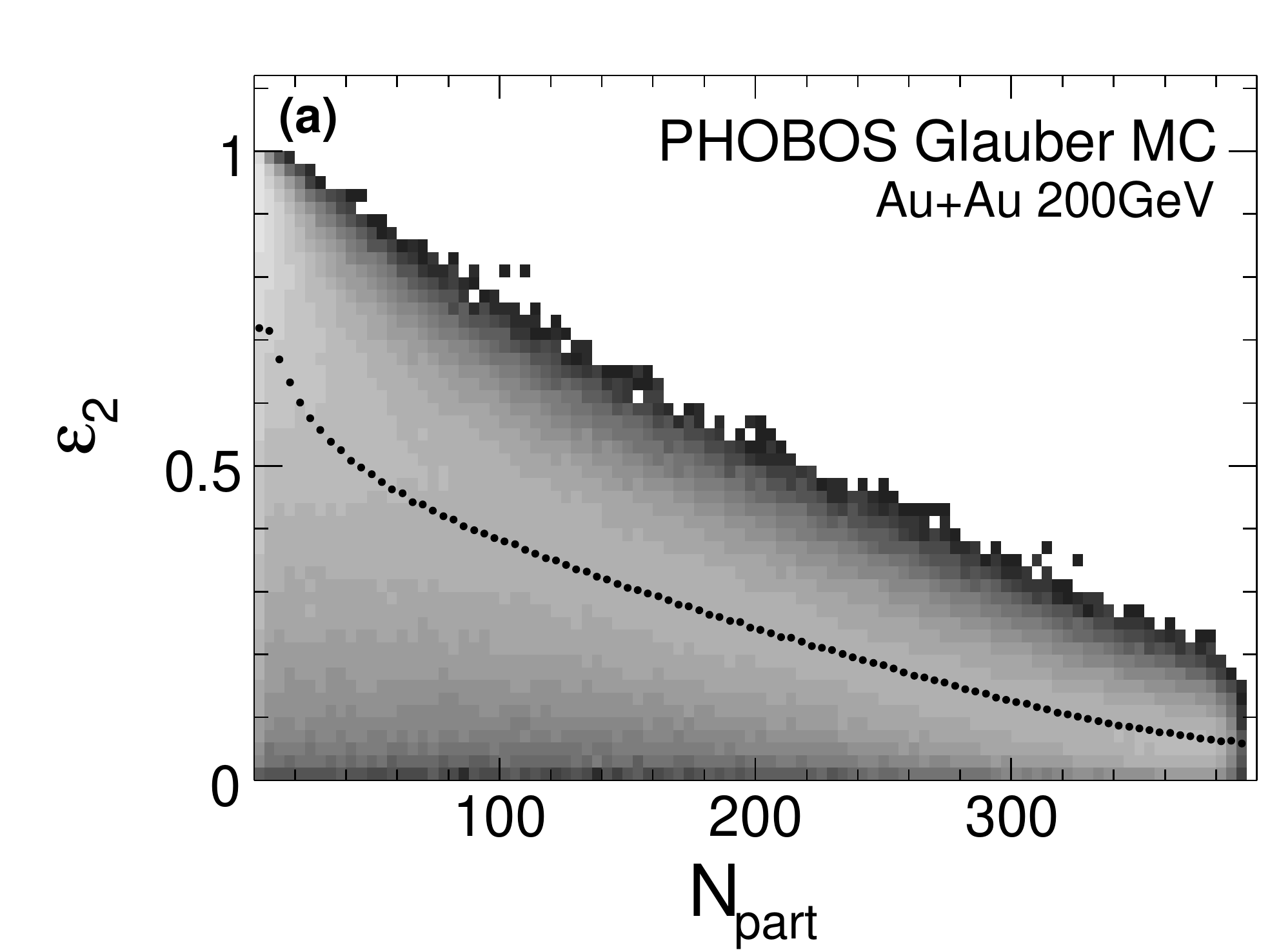}    
    \label{fig:glauecc}
  }
  \subfigure{
    \includegraphics[width=0.45\textwidth]{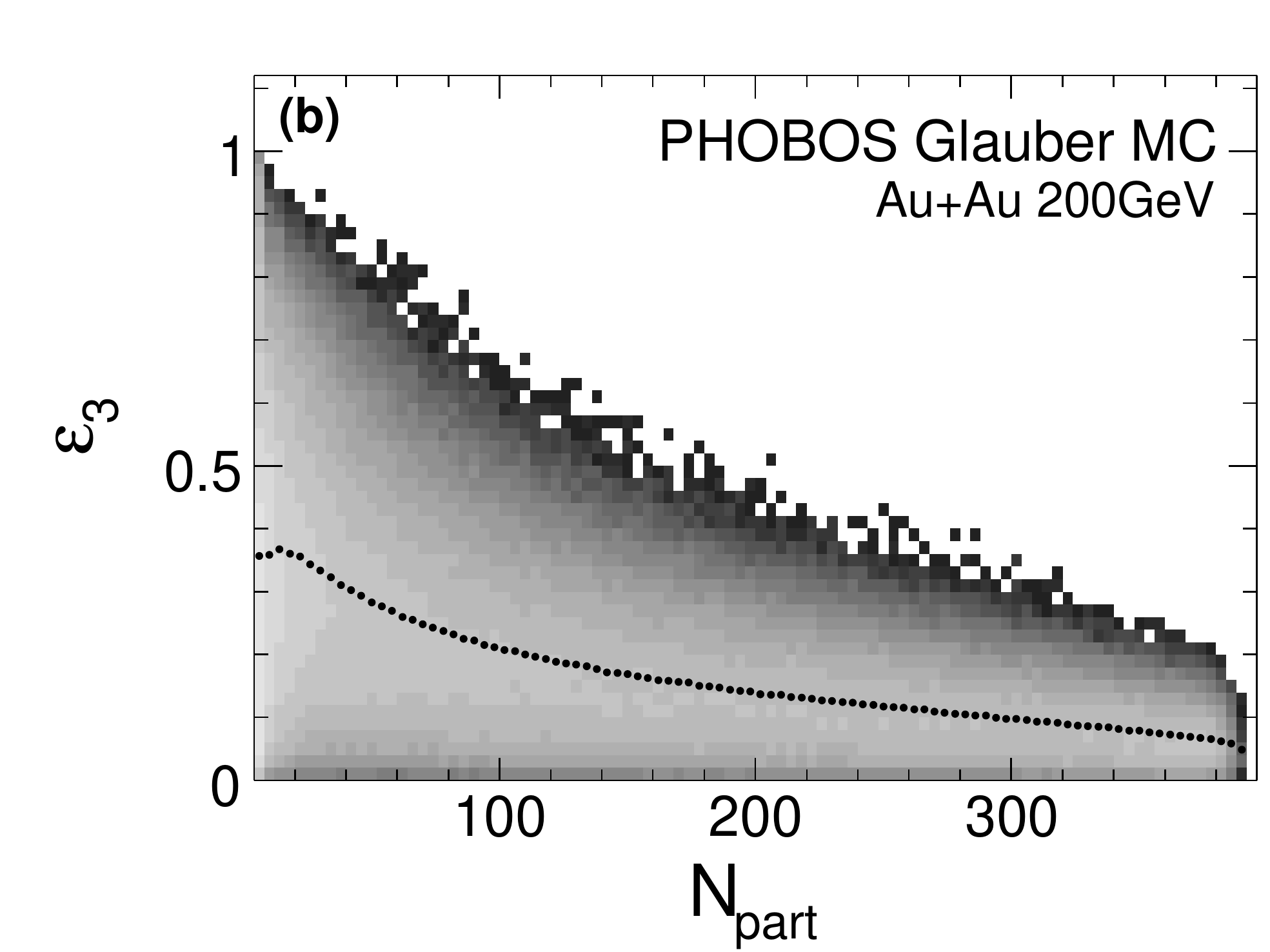}
    \label{fig:glautria}
  }
  \caption{Distribution of \subref{fig:glauecc} eccentricity, $\ecc$,
    and \subref{fig:glautria} triangularity, $\tria$, as a
    function of number of participating nucleons, $\Npart$, in
    \snntwo\ Au+Au collisions.}
  \label{fig:glau1}
\end{figure*}

\begin{figure}[b]
  \includegraphics[width=0.45\textwidth]{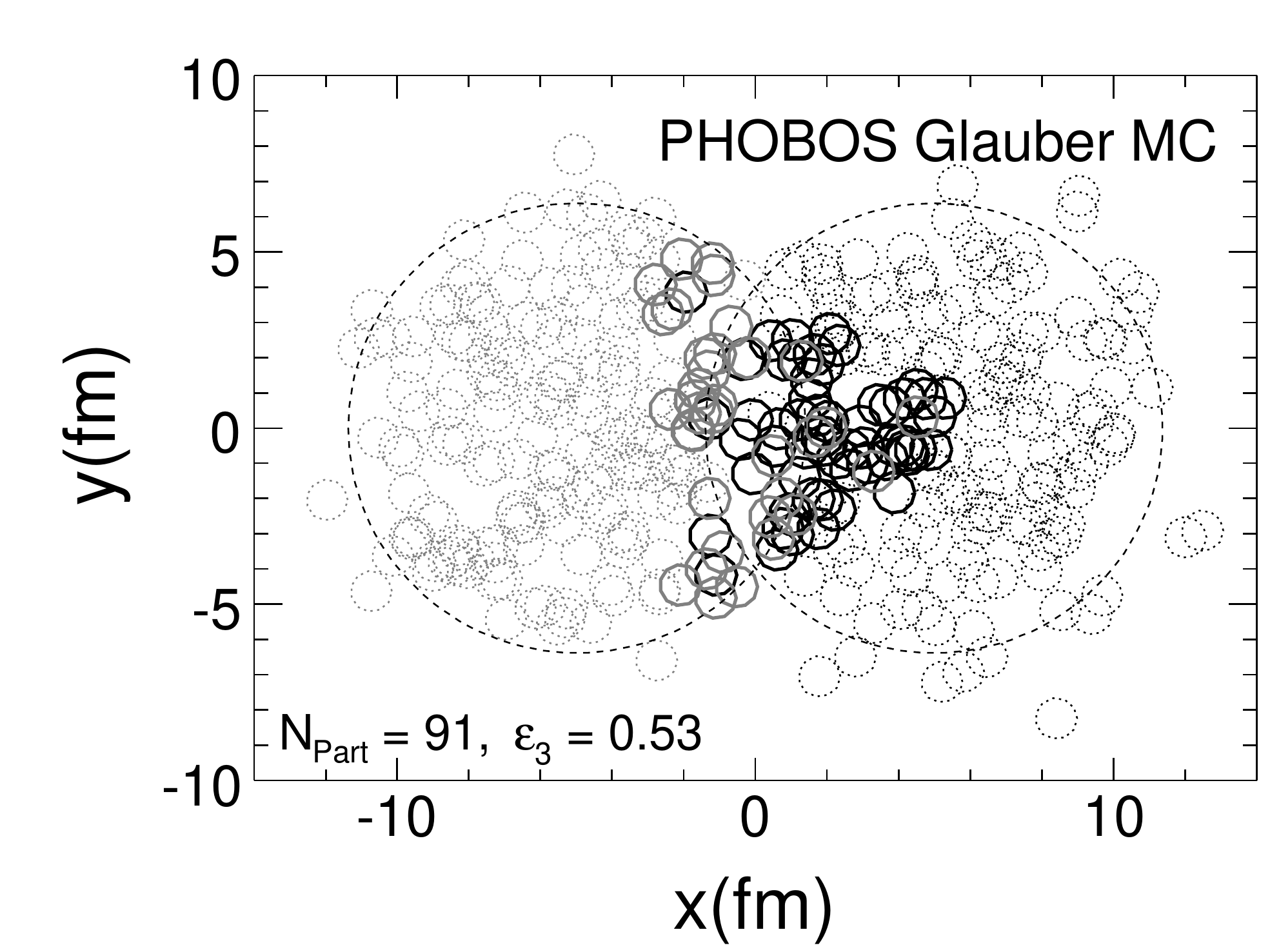}
  \caption{Distribution of nucleons on the transverse plane for a
    \snntwo\ Au+Au collision event with $\tria$=0.53 from
    Glauber Monte Carlo.
    The nucleons in the two nuclei are shown in gray and
    black. Wounded nucleons (participants) are indicated as solid
    circles, while spectators are dotted circles.}
\label{fig:glau2}
\end{figure}

\section{Fourier decomposition of azimuthal correlations}

In the existing correlation data, different correlation measures such
as $R(\deta,\dphi)$~\cite{Alver:2008gk},
$N\hat{r}(\deta,\dphi)$~\cite{Abelev:2008un} and
$1/N_{\text{trig}}\der N / \der
\dphi(\deta,\dphi)$~\cite{Alver:2009id} have been used to study
different sources of particle correlations. The azimuthal projection
of all of these correlation functions have the form
\begin{equation}
  C(\dphi) =  A \frac{\der N^{\text{pairs}}}{\der \dphi} + B, 
\label{eq:Cdphi}
\end{equation}
where the scale factor $A$ and offset $B$ depend on the definition of
the correlation function as well as the \prap\ range of the
projection~\cite{Alver:2009id}.  Examples of long range azimuthal
correlation distributions are shown in \fig{fig:corrs1d}
for mid-central Au+Au collisions  with different trigger and associated
particle $p_T$ selections  obtained by projecting the
two-dimensional correlation functions onto the $\dphi$ axis at \prap\
separations of $1.2<\deta<1.9$ for STAR data~\cite{Abelev:2008un} 
 and $2<\deta<4$ for
PHOBOS data~\cite{Alver:2008gk,Alver:2009id}. The correlation function data used in this study are
available at~\cite{Phobosdata1, Phobosdata2, Stardata1}.  Also shown
in \fig{fig:corrs1d} are the first three Fourier components of the
azimuthal correlations and the residual after these components are
taken out. The data is found to be very well described by the three
Fourier components.
\begin{figure*}[t]
  \centering
  \includegraphics[width=0.9\textwidth]{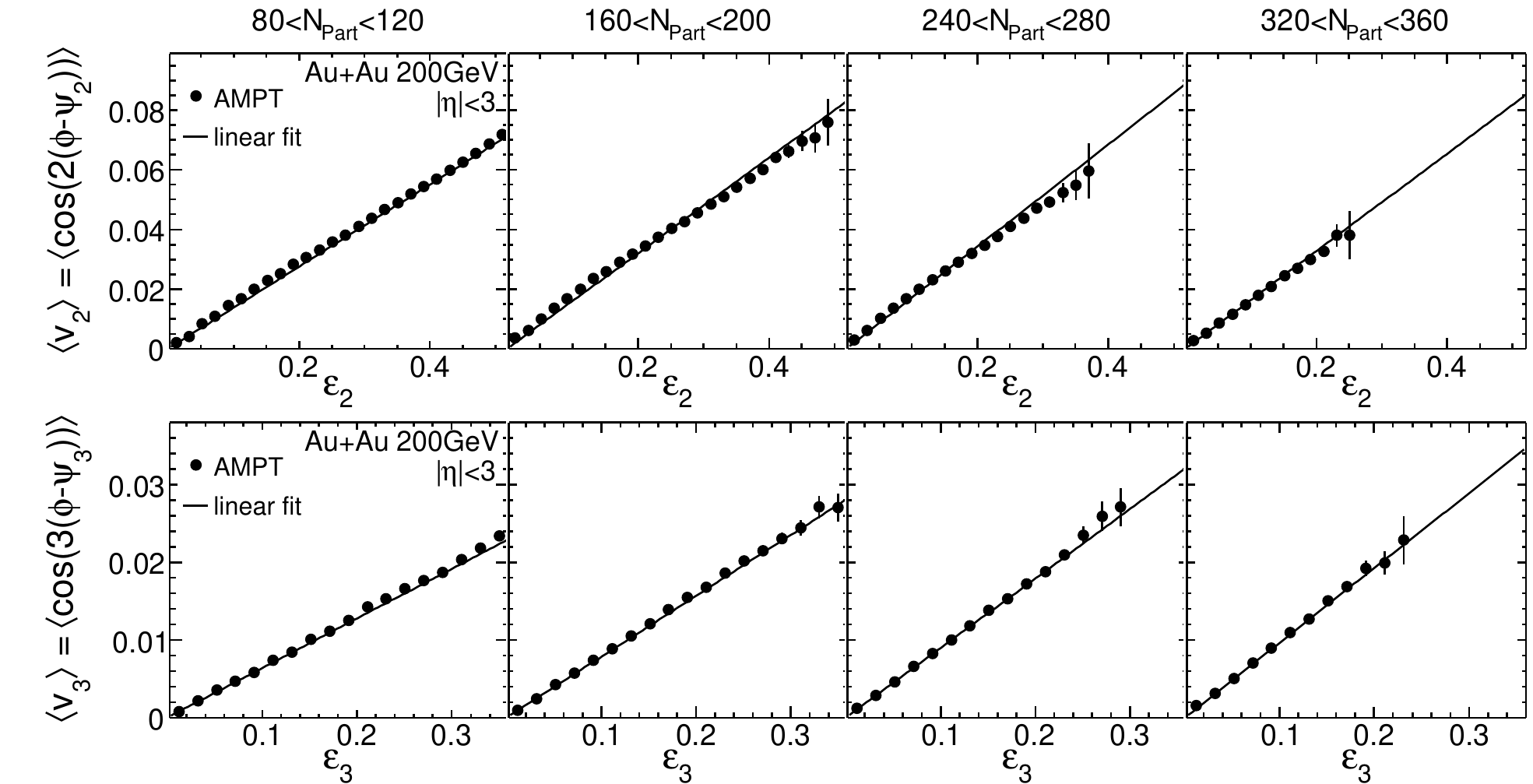}
  \caption{Top: average elliptic flow, $\mean{v_2}$, as a function of
    eccentricity, $\ecc$; bottom: average triangular flow,
    $\mean{v_3}$, as a function of triangularity, $\tria$,
    in \snntwo\ Au+Au collisions from the AMPT model in bins of
    number of participating nucleons. Error bars indicate statistical
    errors.  A linear fit to the data is shown.}
  \label{fig:AMPT_vnVsIni_npartslices_wrtpsi}
\end{figure*}

\section{Participant triangularity and triangular flow}

It is useful to recall that traditional hydrodynamic
calculations start from a smooth matter distribution given by the
transverse overlap of two Woods-Saxon distributions. In such
calculations, elliptic flow is aligned with the orientation of the
reaction plane defined by the impact parameter direction and the beam
axis and by symmetry, no $\V{3}$ component arises in the azimuthal
correlation function. To describe this component in terms of
hydrodynamic flow requires a revised understanding of the initial
collision geometry, taking into account fluctuations in the
nucleon-nucleon collision points from event to event. 
The possible influence of initial geometry fluctuations was used to
explain the surprisingly large values of elliptic flow measured for
central Cu+Cu collision, where the average eccentricity calculated
with respect to the reaction plane angle is small~\cite{Alver:2006wh}.
For a Glauber Monte Carlo event, the minor axis of eccentricity of the
region defined by nucleon-nucleon interaction points does not
necessarily point along the reaction plane vector, but may be tilted.
The ``participant eccentricity''~\cite{Alver:2006wh, Alver:2008zza}
calculated with respect to this tilted axis is found to be finite even
for most central events and significantly larger than the reaction
plane eccentricity for the smaller Cu+Cu system. Following this idea,
event-by-event elliptic flow fluctuations have been measured and found
to be consistent with the expected fluctuations in the initial state
geometry with the new definition of eccentricity~\cite{Alver:2010rt}.
In this paper, we use this method of quantifying the initial
anisotropy exclusively.

Mathematically, the participant eccentricity is given as
\begin{equation}
\ecc = \frac{\sqrt{(\sigma_{y}^2-\sigma_{x}^2)^2 + 4(\sigma_{xy})^2}}{\sigma_{y}^2+\sigma_{x}^2},
\end{equation}
where $\sigma_{x}^2$, $\sigma_{y}^2$ and $\sigma_{xy}$, are the
event-by-event (co)variances of the participant nucleon distributions
along the transverse directions $x$ and $y$~\cite{Alver:2006wh}. If
the coordinate system is shifted to the center of mass of the
participating nucleons such that $\mean{x}=\mean{y}=0$, it can be
shown that the definition of eccentricity is equivalent to
\begin{equation}
  \ecc = \frac{\sqrt{\mean{r^2\cos(2\phi_{\text{part}})}^2 + \mean{r^2\sin(2\phi_{\text{part}})}^2}}{\mean{r^2}}
\label{eq:ecc}
\end{equation}
in this shifted frame, where $r$ and $\phi_{\text{part}}$ are the
polar coordinate positions of participating nucleons.  The minor axis
of the ellipse defined by this region is given as
\begin{equation}
  \psi_{2}=\frac{\atantwo\left(\mean{r^2\sin(2\phi_{\text{part}})},\mean{r^2\cos(2\phi_{\text{part}})}\right)
+\pi}{2}.
\label{eq:phiecc}
\end{equation}
Since the pressure gradients are largest along $\psi_{2}$, the
collective flow is expected to be the strongest in this direction. The
definition of $v_2$ has conceptually changed to refer to the second
Fourier coefficient of particle distribution with respect to
$\psi_{2}$ rather than the reaction plane
\begin{equation}
v_2 = \mean{\cos(2(\phi-\psi_2))}.
\label{eq:v2}
\end{equation}
This change has not impacted the experimental definition since the
directions of the reaction plane angle or $\psi_{2}$ are not a priori
known.

Drawing an analogy to eccentricity and elliptic flow, the initial and
final triangular anisotropies can be quantified as participant
triangularity, $\tria$, and triangular flow, $v_3$, respectively:
\begin{eqnarray}
  \tria &\equiv& \frac{\sqrt{\mean{r^2\cos(3\phi_{\text{part}})}^2 + \mean{r^2\sin(3\phi_{\text{part}})}^2}}{\mean{r^2}} \label{eq:tria} \\
v_3 &\equiv& \mean{\cos(3(\phi-\psi_3))}
\label{eq:v3}
\end{eqnarray}
 where $\psi_3$ is the minor axis of participant triangularity given by
 \begin{equation}
   \psi_{3}=\frac{\atantwo\left(\mean{r^2\sin(3\phi_{\text{part}})},\mean{r^2\cos(3\phi_{\text{part}})}\right)
     +\pi}{3}.
\label{eq:phitria}
\end{equation}
It is important to note that the minor axis of triangularity is found
to be uncorrelated with the reaction plane angle and the minor axis of
eccentricity in Glauber Monte Carlo calculations. This implies that
the average triangularity calculated with respect to the reaction
plane angle or $\psi_2$ is zero. The participant triangularity defined
in \eq{eq:tria}, however, is calculated with respect to $\psi_3$ and
is always finite.

\begin{figure*}[t]
  \centering
  \subfigure{
    \includegraphics[width=0.45\textwidth]{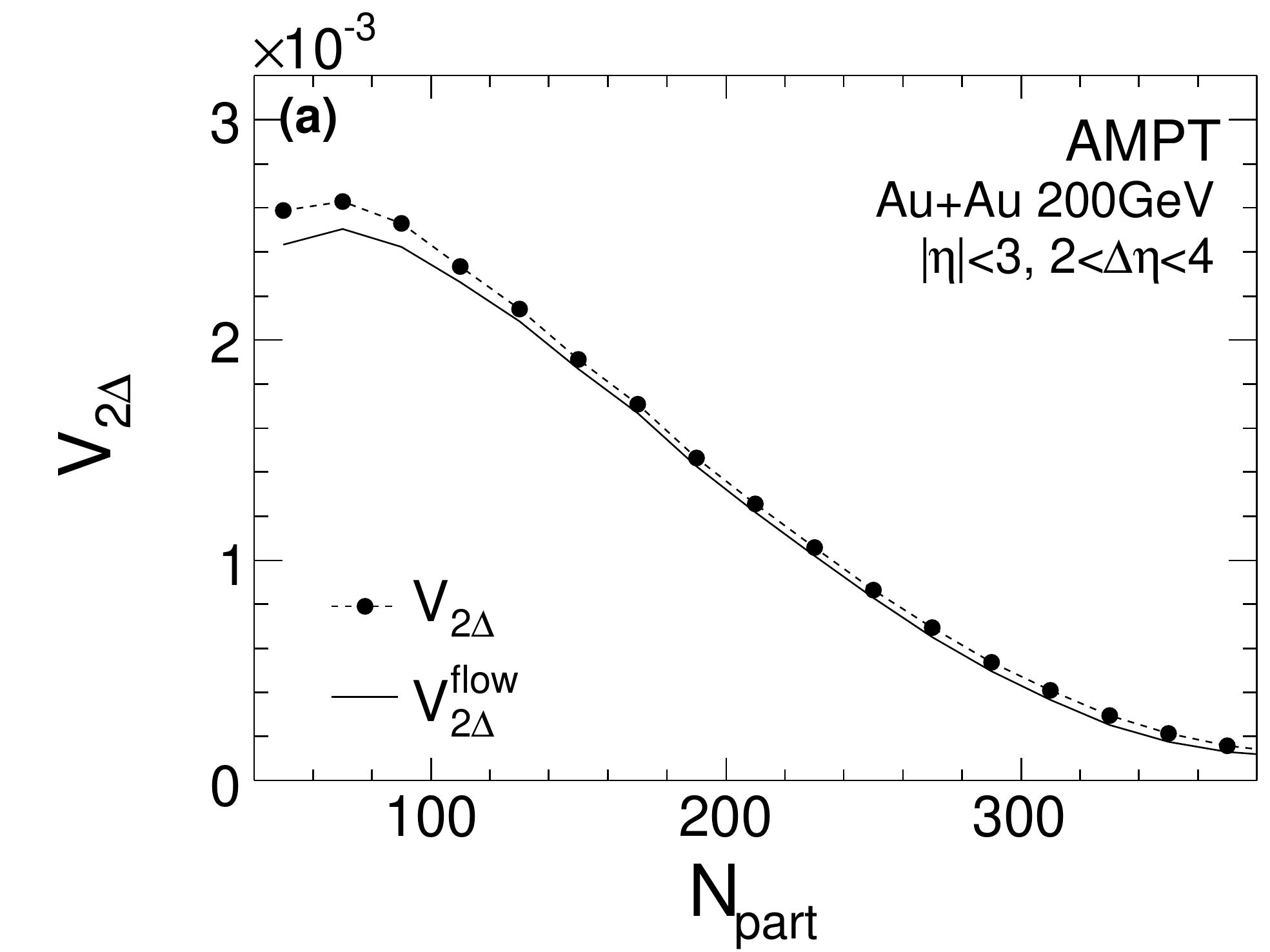}
    \label{fig:AMPT_meanv22andv2_2vsNpart_2_1}
  }
  \subfigure{
    \includegraphics[width=0.45\textwidth]{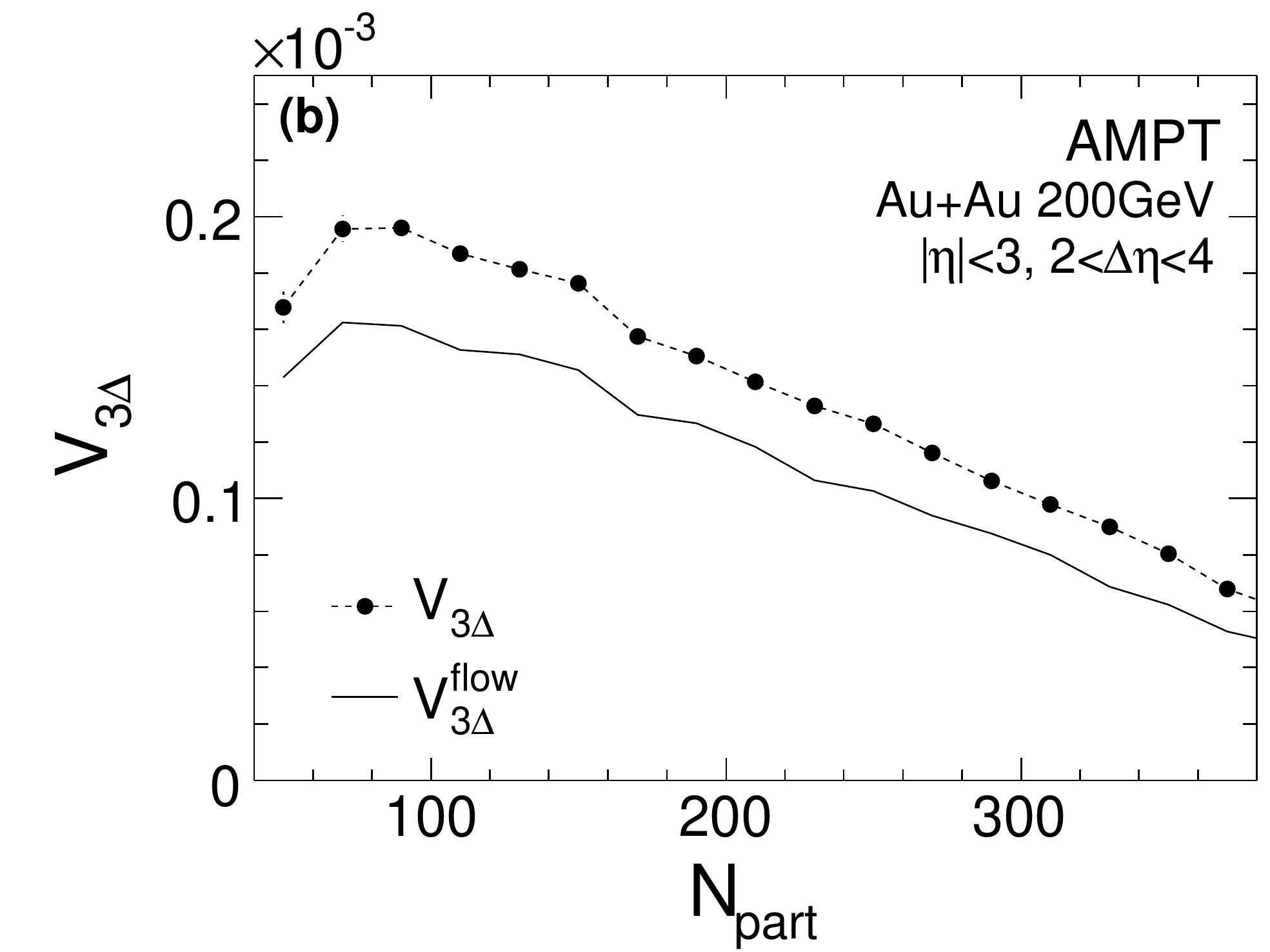}
    \label{fig:AMPT_meanv32andv3_2vsNpart_2_1}
  }
  \caption{Dashed lines show
    \subref{fig:AMPT_meanv22andv2_2vsNpart_2_1} second Fourier
    coefficient, $\V{2}$, and
    \subref{fig:AMPT_meanv32andv3_2vsNpart_2_1} third Fourier
    coefficient, $\V{3}$, of azimuthal correlations as a function of
    number of participating nucleons, $\Npart$, in \snntwo\ Au+Au
    collisions from the AMPT model. Solid lines show the contribution to
    these coefficients from flow calculated with respect to the minor
    axis of (a) eccentricity and (b) triangularity.}
  \label{fig:meansquare}
\end{figure*}

The distributions of eccentricity and triangularity calculated with
the PHOBOS Glauber Monte Carlo implementation~\cite{Alver:2008aq} for
Au+Au events at \snntwo\ are shown in \fig{fig:glau1}. The value of
triangularity is observed to fluctuate event-by-event and have an
average magnitude of the same order as eccentricity.  Transverse
distribution of nucleons for a sample Monte Carlo event with a high
value of triangularity is shown in \fig{fig:glau2}.  A clear
triangular anisotropy can be seen in the region defined by the
participating nucleons. 

\section{Triangular flow in the AMPT model}

To assess the connection between triangularity and the ridge and broad
away-side features in two-particle correlations, we study elliptic and
triangular flow in the AMPT model.  AMPT is a hybrid model which
consists of four main components: initial conditions, parton cascade,
string fragmentation, and A Relativistic Transport Model for
hadrons. The model successfully describes main features of the
dependence of elliptic flow on centrality and transverse
momentum~\cite{Lin:2004en}. Ridge and broad away-side features in
two-particle correlations are also observed in the AMPT
model~\cite{Ma:2006fm,Ma:2008nd}.  Furthermore, the dependence of
quantitative observables such as away-side RMS width and away-side
splitting parameter $D$ on transverse momentum and reaction plane in
AMPT reproduces the experimental results successfully, where a
ZYAM-based elliptic flow subtraction is applied to both the data and
the model~\cite{Zhang:2007qx,Li:2009ti}.

The initial conditions of AMPT are obtained from Heavy Ion Jet
Interaction Generator (HIJING)~\cite{Gyulassy:1994ew}. HIJING uses a
Glauber Model implementation that is similar to the PHOBOS
implementation to determine positions of participating nucleons. It is
possible to calculate the values of $\ecc$, $\psi_2$, $\tria$ and
$\psi_3$ event-by-event from the positions of these nucleons [see
Equations~\ref{eq:ecc}, \ref{eq:phiecc}, \ref{eq:tria}
and~\ref{eq:phitria}]. Next, we calculate the magnitudes of elliptic
and triangular flow with respect to $\psi_2$ and $\psi_3$ respectively
as defined in \eqs{eq:v2}{eq:v3}.

The average value of elliptic flow, $v_2$, and triangular flow, $v_3$,
for particles in the \prap\ range \mbox{$\abs{\eta}\!<\!3$} in
\snntwo\ Au+Au collisions from AMPT are shown as a function of $\ecc$
and $\tria$ in \fig{fig:AMPT_vnVsIni_npartslices_wrtpsi} for different
ranges of number of participating nucleons. As previously expected,
the magnitude of $v_2$ is found to be proportional to $\ecc$.  We
observe that a similar linear relation is also present between
triangular flow and triangularity.

\begin{figure*}[t]
  \centering
  \subfigure{
    \includegraphics[width=0.45\textwidth]{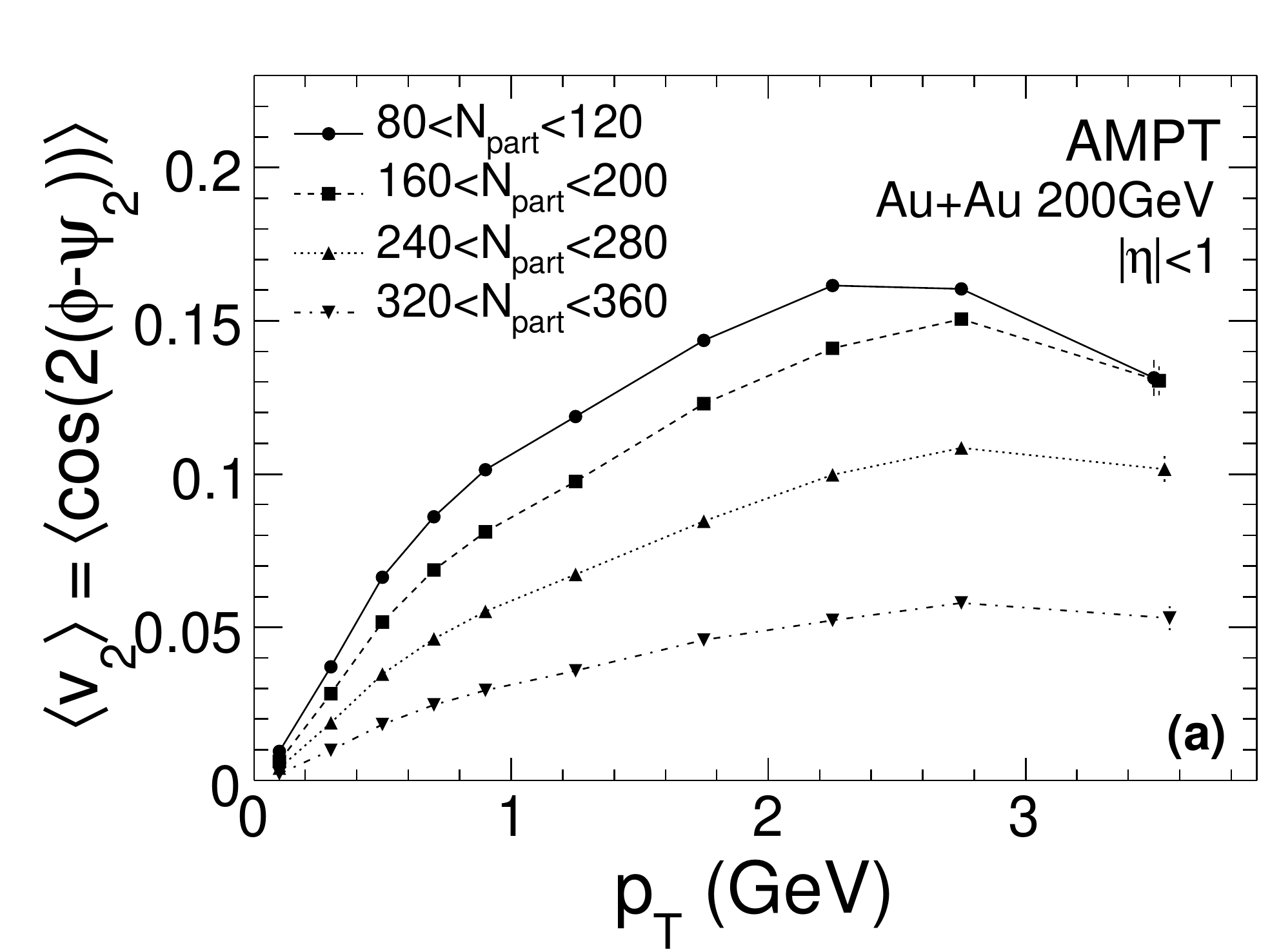}
    \label{fig:AMPTv2vspt}
  }
  \subfigure{
    \includegraphics[width=0.45\textwidth]{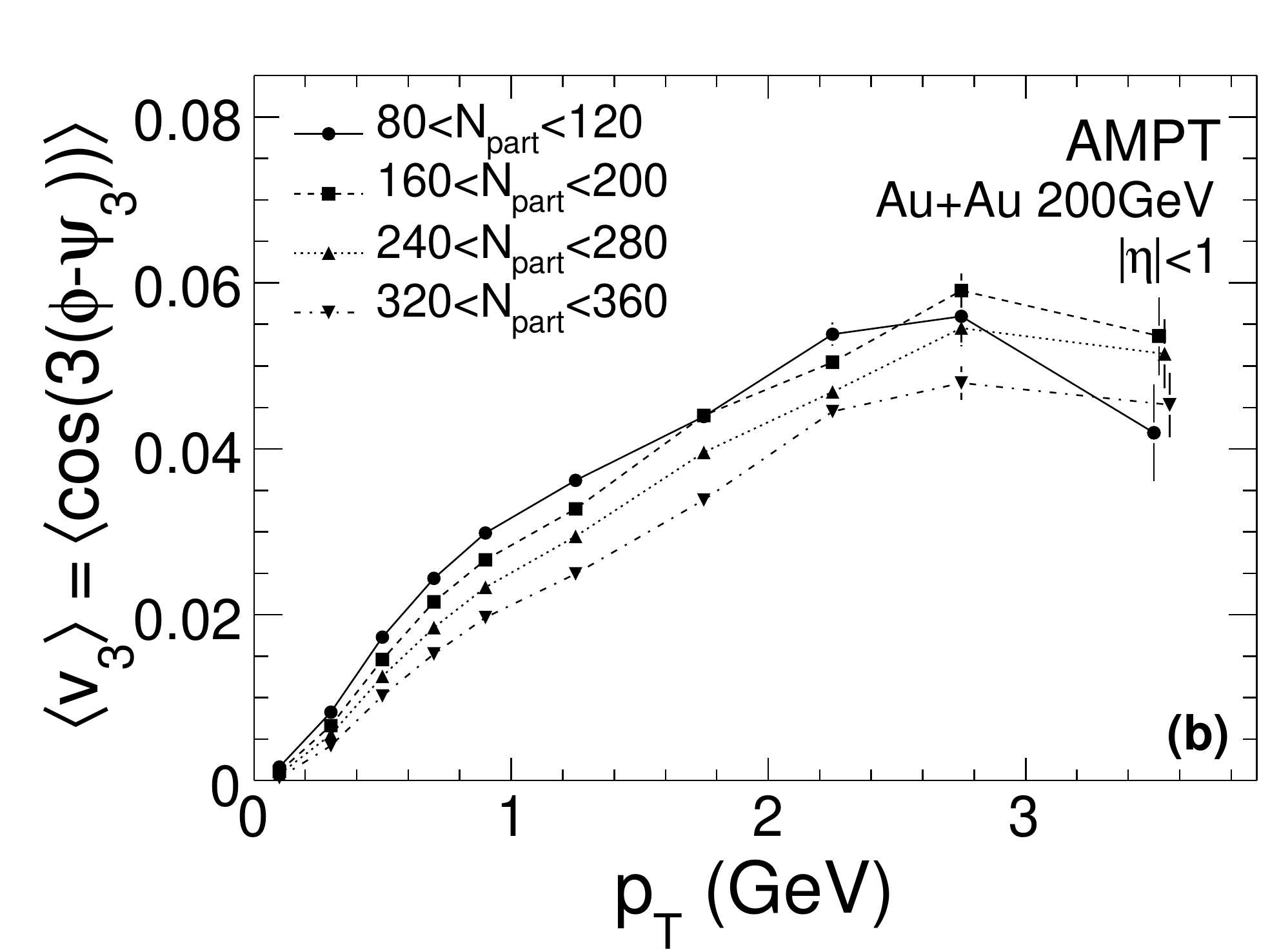}
    \label{fig:AMPTv3vspt}
  }
  \caption{\subref{fig:AMPTv2vspt} Elliptic flow, $v_2$, and
    \subref{fig:AMPTv3vspt} triangular flow, $v_3$, as a function of
    transverse momentum, $\pt$, in bins of number of participating
    nucleons, $\Npart$, for particles at mid-rapidity ($\abs{\eta}<1$)
    in \snntwo\ Au+Au collisions from the AMPT model. Error bars indicate
    statistical errors.}
  \label{fig:pt}
\end{figure*}

\begin{figure}[b]
\includegraphics[width=0.45\textwidth]{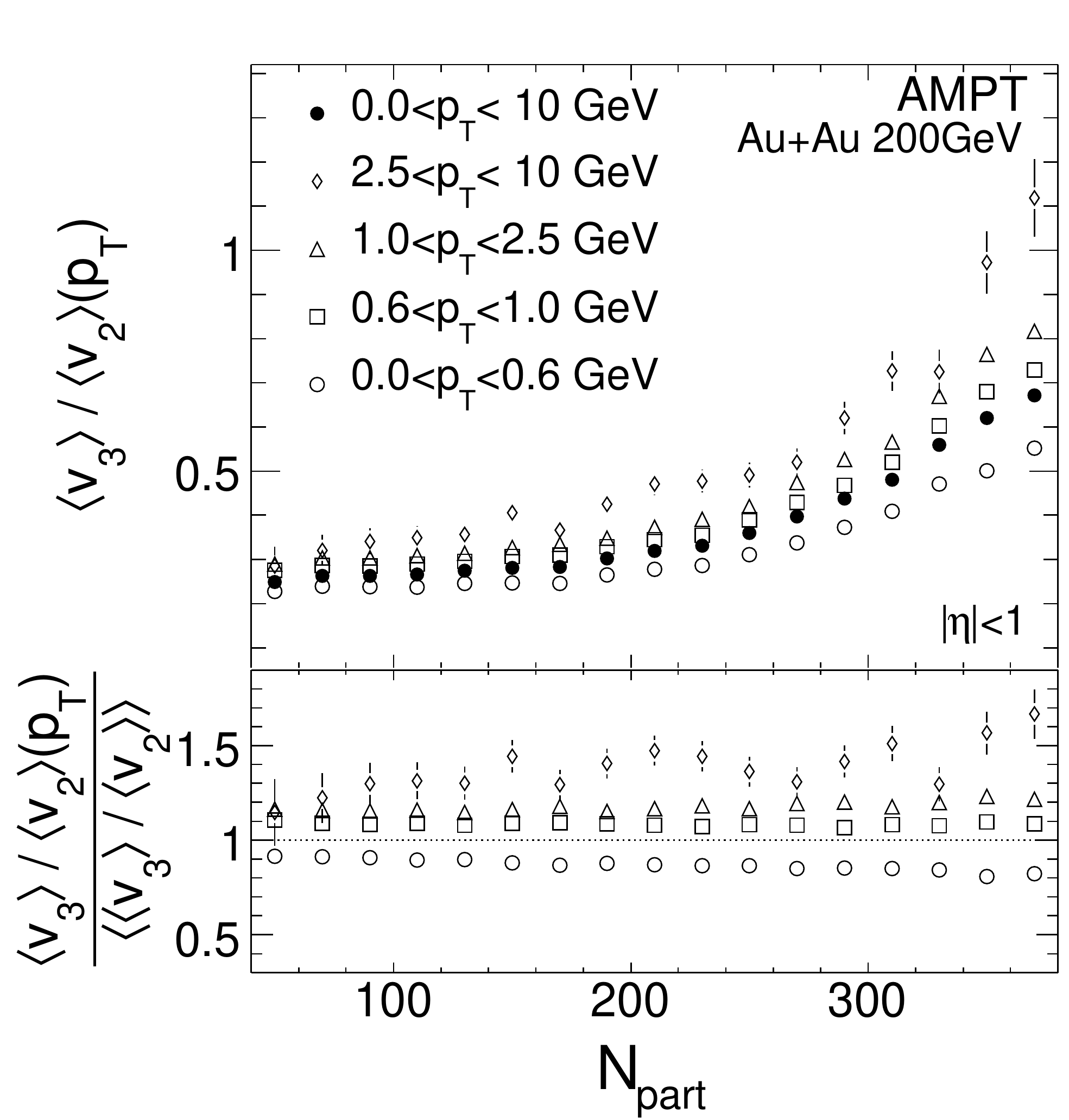}
\caption{Top: the ratio of triangular flow to elliptic flow,
  $\mean{v_3}/\mean{v_2}$, as a function of number of participating
  nucleons, $\Npart$, for particles at mid-rapidity ($\abs{\eta}<1$)
  in \snntwo\ Au+Au collisions from the AMPT model. Open points show
  different transverse momentum bins and the filled points show the
  average over all transverse momentum bins. Bottom: the ratio of
  different $\pt$ bins to the average value. Error bars indicate
  statistical errors.}
\label{fig:AMPTv3overv2ptnpart}
\end{figure}

After establishing that triangular anisotropy in initial collision
geometry leads to a triangular anisotropy in particle production, we
investigate the contribution of triangular flow to the observed ridge
and broad away-side features in two-particle azimuthal correlations.
For a given \prap\ window, the Fourier coefficients of two-particle
azimuthal correlations, $\V{n}$, can be calculated in AMPT by averaging
$\cos(n\dphi)$ over all particle pairs. Contributions from elliptic
(triangular) flow is present in the second (third) Fourier coefficient
of $\dphi$ distribution since
\begin{multline}
\int \frac{1}{4\pi^2}  \left\{1+2v_n\cos(n\phi)\right\} \\
\times \left\{1+2v_n\cos(n(\phi+\dphi))\right\} \der \phi \qquad \qquad \\
  = \frac{1}{2\pi} \left\{1+2v_n^2 \cos(n\dphi)\right\}.
\label{eq:vnflow1}
\end{multline}
For a given \prap\  window, this contribution can be calculated
from average elliptic (triangular) flow values as
\begin{multline}
 \V{n}^{\text{flow}}=\frac{\mean{\varepsilon_n^2}}{\mean{\varepsilon_n}^2} \\
 \times \frac{\int\frac{\der N}{\der \eta}(\eta_{1}) \frac{\der N}{\der \eta}(\eta_{2}) \mean{v_{n}(\eta_1)} \mean{v_{n}(\eta_2)} \der \eta_1 \der \eta_2}{\int \frac{\der N}{\der \eta}(\eta_{1}) \frac{\der N}{\der \eta}(\eta_{2}) \der \eta_1 \der \eta_2} 
\label{eq:vnflow}
\end{multline}
where $n\! = \! 2$ ($n\! = \! 3$) and the integration is over the
\prap\ range of particle pairs. The average single-particle
distribution coefficients, $\mean{v_{n}(\eta)}$, are used in this
calculation to avoid contributions from non-flow correlations which may
be present if the two-particle distributions, $v_{n}(\eta_1)\times
v_{n}(\eta_2)$, are calculated event by event. The ratio
$\mean{\varepsilon_n^2}/\mean{\varepsilon_n}^2$ accounts for the difference
between $\mean{v_n(\eta_1) \times v_n(\eta_2)}$ and
$\mean{v_n(\eta_1)} \times \mean{v_n(\eta_2)}$ expected from initial
geometry fluctuations.

We have calculated the magnitude of the second and third Fourier
components of two-particle azimuthal correlations and expected
contributions to these components from elliptic and triangular flow
for particle pairs in \snntwo\ Au+Au collisions from AMPT within the
\prap\  range $\abs{\eta}<3$ and $2<\deta<4$. The results are
presented in \fig{fig:meansquare} as a function of number of
participating nucleons. More than 80\% of the third Fourier
coefficient of azimuthal correlations can be accounted for by
triangular flow with respect to the minor axis of triangularity.  The
difference between $\V{3}$ and $\V{3}^{\text{flow}}$ may be due to
two different effects: There might be contributions from correlations
other than triangular flow to $\V{3}$ or the angle with respect to
which the global triangular anisotropy develops might not be given
precisely by the minor axis of triangularity calculated from
positions of participant nucleons, i.e.\
$v_3=\mean{(\cos(3(\phi-\psi_3)}$ might be an underestimate for the
magnitude of triangular flow. More detailed studies are needed to
distinguish between these two effects.

We have also studied the magnitudes of elliptic and triangular flow
more differentially as a function of transverse momentum and number of
participating nucleons in the AMPT model. \Fig{fig:pt} shows the results
as a function of transverse momentum for particles at mid-rapidity
($\abs{\eta}<1$) for different ranges of number of participating
nucleons. The dependence of triangular flow on transverse momentum is
observed to show similar gross features as elliptic flow. A more
detailed comparison can be made by taking the ratio of triangular to
elliptic flow, shown in \fig{fig:AMPTv3overv2ptnpart} as a function of
number of participating nucleons for different ranges of transverse
momentum. The relative strength of triangular flow is observed to
increase with centrality and transverse momentum. This observation is
qualitatively consistent with the trends in experimentally measured
ridge yield~\cite{Alver:2009id}.

\section{Triangular flow in experimental data}
\begin{figure*}[t]
  \centering
  \subfigure{
    \includegraphics[width=0.45\textwidth]{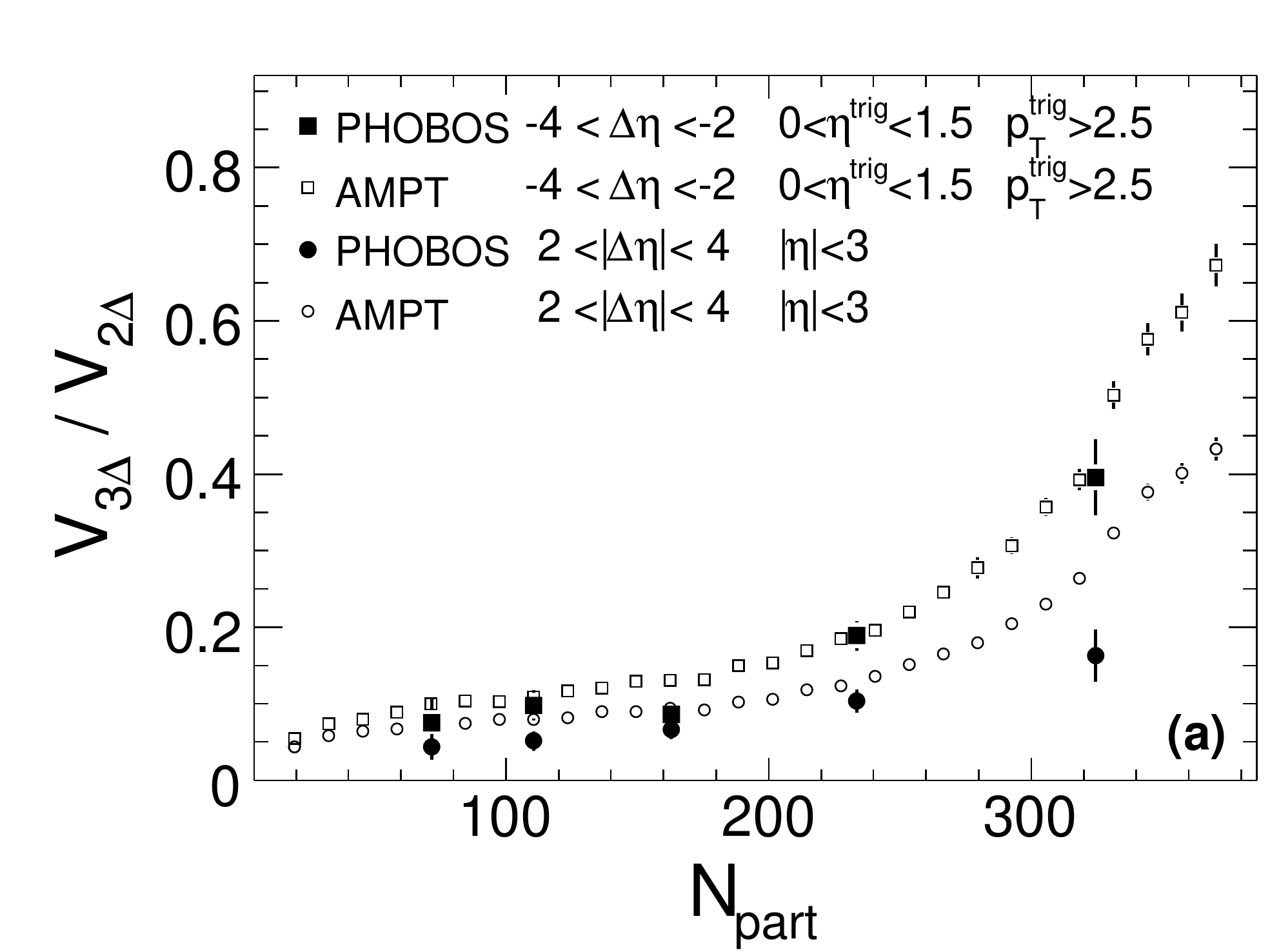}
    \label{fig:V2OverV3VsNpart_a}
  }
  \subfigure{
    \includegraphics[width=0.45\textwidth]{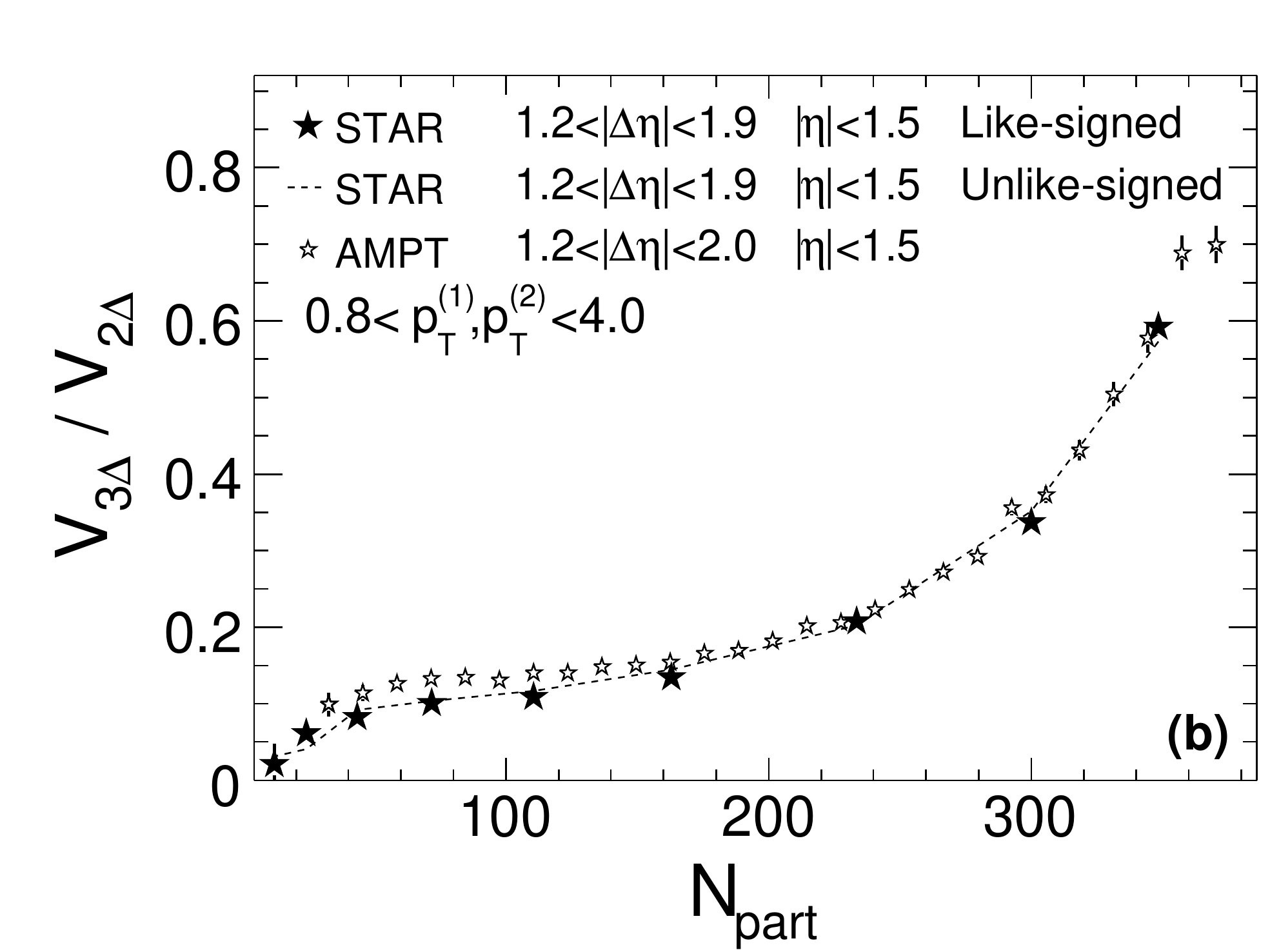}
    \label{fig:V2OverV3VsNpart_b}
  }
  \caption{The ratio of the third to second Fourier coefficients of
    azimuthal correlations, $\V{3}/\V{2}$, as a function of number of
    participating nucleons, $\Npart$, for Au+Au collisions at \snntwo.
    Filled points show values derived from \subref{fig:V2OverV3VsNpart_a}
    PHOBOS~\cite{Alver:2009id, Alver:2008gk} and
    \subref{fig:V2OverV3VsNpart_b} STAR~\cite{Abelev:2008un}
    data. \Prap\  and transverse momentum
    ranges and charge selection of particle pairs for different
    measurements are indicated on the figures.  Open points show
    results from the AMPT model for similar selection of \prap\ 
    and transverse momentum to the available data.  Error bars
    indicate statistical errors for AMPT and combined statistical and
    systematic errors for the experimental data.}
  \label{fig:V2OverV3VsNpart}
\end{figure*}
While AMPT reproduces the expected proportionality of $v_2$ and $\ecc$,
the absolute magnitude of $v_2$ is underestimated compared to data and 
hydrodynamic calculations. To allow a comparison of the $\V{3}$ calculations
to data, we therefore use the ratio of the third and second Fourier
coefficients. For data, this ratio is given by
\begin{equation}
\frac{\V{3}}{\V{2}} = \frac{\int C(\dphi) \cos(3\dphi) \der \dphi}{\int C(\dphi) \cos(2\dphi) \der \dphi}.
\end{equation}
The factors $A$ and $B$ in \eq{eq:Cdphi} cancel out in this ratio.
Results for PHOBOS~\cite{Alver:2009id, Alver:2008gk} and
STAR~\cite{Abelev:2008un} measurements are plotted as a function of
number of participating nucleons in
Figures~\ref{fig:V2OverV3VsNpart_a} and
\subref{fig:V2OverV3VsNpart_b}, respectively.  It is observed that
$\V{3}/\V{2}$ increases with centrality and with the transverse momentum
of the particles. 

Also shown in \fig{fig:V2OverV3VsNpart} is the magnitude of
$\V{3}/\V{2}$ in the AMPT model with similar $\eta$, $\deta$ and $\pt$
selections to the available experimental data. The calculations from
the model show a qualitative agreement with the data in term of the
dependence of $\V{3}/\V{2}$ on the \prap\ region, particle
momenta and centrality. Since the $\V{3}$ component of two-particle
correlations in the model is known to be mostly due to the triangular
anisotropy in the initial collision geometry, this observation
suggests that triangular flow may play an important role in
understanding the ridge and broad away-side structures in data.

A closer look at the properties of the ridge and broad away-side is
possible via studies of three particle correlations. Triangular flow
predicts a very distinct signature in three particle correlation
measurements. Two recent publications by the STAR experiment present
results on correlations in $\dphi_1$-$\dphi_2$ space for
$\abs{\eta}<1$~\cite{:2008nda} and in $\deta_1$-$\deta_2$ space for
$\abs{\dphi}<0.7$~\cite{Abelev:2009jv}. In $\dphi_1$-$\dphi_2$ space,
off diagonal away-side correlations have been observed (e.g.\ first
associated particle at $\dphi_1\approx 120^{\circ}$ and second
associated particle at $\dphi_2\approx -120^{\circ}$) consistent with
expectations from triangular flow. In $\deta_1$-$\deta_2$ space, no
correlation structure between the two associated ridge particles was
detected, also consistent with triangular flow.


\section{Summary}

We have introduced the concepts of participant
triangularity and triangular flow, which quantify the triangular
anisotropy in the initial and final states of heavy-ion collisions.
It has been shown that inclusive and triggered two-particle azimuthal
correlations at large $\deta$ in heavy-ion collisions are well
described by the first three Fourier components.  It has been
demonstrated that event-by-event fluctuations lead to a finite
triangularity value in Glauber Monte Carlo events and that this
triangular anisotropy in the initial geometry leads to a triangular
anisotropy in particle production in the AMPT model. The third Fourier
coefficient of azimuthal correlations at large \prap\ separations have
been found to be dominated by triangular flow in the model. We have
studied the ratio of the third and second Fourier coefficients of
azimuthal correlations in experimental data and the AMPT model as a
function of centrality and \prap\ and momentum ranges of particle pairs. A
qualitative agreement between data and model has been observed. This
suggests that the ridge and broad away-side features observed in
two-particle correlation measurements in Au+Au collisions contain a
significant, and perhaps dominant, contribution from triangular flow.
Our findings support previous evidence from measurements of the system
size dependence of elliptic flow and elliptic flow fluctuations on the
importance of geometric fluctuations in the initial collision
region. Detailed studies of triangular flow can shed new light on the
initial conditions and the collective expansion of the matter created
in heavy-ion collisions.

The authors acknowledge fruitful discussions with Wei Li, Constantin
Loizides, Peter Steinberg and Edward Wenger. This work was supported
by U.S. DOE grant DE-FG02-94ER40818.

\bibliographystyle{apsrev} 
\bibliography{v3paper}

\begin{thebibliography}{53}
\expandafter\ifx\csname natexlab\endcsname\relax\def\natexlab#1{#1}\fi
\expandafter\ifx\csname bibnamefont\endcsname\relax
  \def\bibnamefont#1{#1}\fi
\expandafter\ifx\csname bibfnamefont\endcsname\relax
  \def\bibfnamefont#1{#1}\fi
\expandafter\ifx\csname citenamefont\endcsname\relax
  \def\citenamefont#1{#1}\fi
\expandafter\ifx\csname url\endcsname\relax
  \def\url#1{\texttt{#1}}\fi
\expandafter\ifx\csname urlprefix\endcsname\relax\def\urlprefix{URL }\fi
\providecommand{\bibinfo}[2]{#2}
\providecommand{\eprint}[2][]{\url{#2}}

\bibitem[{\citenamefont{Ackermann et~al.}(2001)}]{Ackermann:2000tr}
\bibinfo{author}{\bibfnamefont{K.~H.} \bibnamefont{Ackermann}}
  \bibnamefont{et~al.} (\bibinfo{collaboration}{STAR}), \bibinfo{journal}{Phys.
  Rev. Lett.} \textbf{\bibinfo{volume}{86}}, \bibinfo{pages}{402}
  (\bibinfo{year}{2001}).

\bibitem[{\citenamefont{Adcox et~al.}(2002)}]{Adcox:2002ms}
\bibinfo{author}{\bibfnamefont{K.}~\bibnamefont{Adcox}} \bibnamefont{et~al.}
  (\bibinfo{collaboration}{PHENIX}), \bibinfo{journal}{Phys. Rev. Lett.}
  \textbf{\bibinfo{volume}{89}}, \bibinfo{pages}{212301}
  (\bibinfo{year}{2002}).

\bibitem[{\citenamefont{Adler et~al.}(2003{\natexlab{a}})}]{Adler:2003kt}
\bibinfo{author}{\bibfnamefont{S.~S.} \bibnamefont{Adler}} \bibnamefont{et~al.}
  (\bibinfo{collaboration}{PHENIX}), \bibinfo{journal}{Phys. Rev. Lett.}
  \textbf{\bibinfo{volume}{91}}, \bibinfo{pages}{182301}
  (\bibinfo{year}{2003}{\natexlab{a}}).

\bibitem[{\citenamefont{Adams et~al.}(2004)}]{Adams:2003am}
\bibinfo{author}{\bibfnamefont{J.}~\bibnamefont{Adams}} \bibnamefont{et~al.}
  (\bibinfo{collaboration}{STAR}), \bibinfo{journal}{Phys. Rev. Lett.}
  \textbf{\bibinfo{volume}{92}}, \bibinfo{pages}{052302}
  (\bibinfo{year}{2004}).

\bibitem[{\citenamefont{Adams et~al.}(2005{\natexlab{a}})}]{Adams:2004bi}
\bibinfo{author}{\bibfnamefont{J.}~\bibnamefont{Adams}} \bibnamefont{et~al.}
  (\bibinfo{collaboration}{STAR}), \bibinfo{journal}{Phys. Rev.}
  \textbf{\bibinfo{volume}{C72}}, \bibinfo{pages}{014904}
  (\bibinfo{year}{2005}{\natexlab{a}}).

\bibitem[{\citenamefont{Back et~al.}(2005{\natexlab{a}})}]{Back:2004zg}
\bibinfo{author}{\bibfnamefont{B.~B.} \bibnamefont{Back}} \bibnamefont{et~al.}
  (\bibinfo{collaboration}{PHOBOS}), \bibinfo{journal}{Phys. Rev. Lett.}
  \textbf{\bibinfo{volume}{94}}, \bibinfo{pages}{122303}
  (\bibinfo{year}{2005}{\natexlab{a}}).

\bibitem[{\citenamefont{Back et~al.}(2005{\natexlab{b}})}]{Back:2004mh}
\bibinfo{author}{\bibfnamefont{B.~B.} \bibnamefont{Back}} \bibnamefont{et~al.}
  (\bibinfo{collaboration}{PHOBOS}), \bibinfo{journal}{Phys. Rev.}
  \textbf{\bibinfo{volume}{C72}}, \bibinfo{pages}{051901}
  (\bibinfo{year}{2005}{\natexlab{b}}).

\bibitem[{\citenamefont{Alver et~al.}(2007)}]{Alver:2006wh}
\bibinfo{author}{\bibfnamefont{B.}~\bibnamefont{Alver}} \bibnamefont{et~al.}
  (\bibinfo{collaboration}{PHOBOS}), \bibinfo{journal}{Phys. Rev. Lett.}
  \textbf{\bibinfo{volume}{98}}, \bibinfo{pages}{242302}
  (\bibinfo{year}{2007}).

\bibitem[{\citenamefont{Adare et~al.}(2007)}]{Adare:2006ti}
\bibinfo{author}{\bibfnamefont{A.}~\bibnamefont{Adare}} \bibnamefont{et~al.}
  (\bibinfo{collaboration}{PHENIX}), \bibinfo{journal}{Phys. Rev. Lett.}
  \textbf{\bibinfo{volume}{98}}, \bibinfo{pages}{162301}
  (\bibinfo{year}{2007}).

\bibitem[{\citenamefont{Adler et~al.}(2003{\natexlab{b}})}]{Adler:2002tq}
\bibinfo{author}{\bibfnamefont{C.}~\bibnamefont{Adler}} \bibnamefont{et~al.}
  (\bibinfo{collaboration}{STAR}), \bibinfo{journal}{Phys. Rev. Lett.}
  \textbf{\bibinfo{volume}{90}}, \bibinfo{pages}{082302}
  (\bibinfo{year}{2003}{\natexlab{b}}).

\bibitem[{\citenamefont{Adams et~al.}(2006{\natexlab{a}})}]{Adams:2006yt}
\bibinfo{author}{\bibfnamefont{J.}~\bibnamefont{Adams}} \bibnamefont{et~al.}
  (\bibinfo{collaboration}{STAR}), \bibinfo{journal}{Phys. Rev. Lett.}
  \textbf{\bibinfo{volume}{97}}, \bibinfo{pages}{162301}
  (\bibinfo{year}{2006}{\natexlab{a}}).

\bibitem[{\citenamefont{Adare et~al.}(2008{\natexlab{a}})}]{:2008cqb}
\bibinfo{author}{\bibfnamefont{A.}~\bibnamefont{Adare}} \bibnamefont{et~al.}
  (\bibinfo{collaboration}{PHENIX}), \bibinfo{journal}{Phys. Rev.}
  \textbf{\bibinfo{volume}{C78}}, \bibinfo{pages}{014901}
  (\bibinfo{year}{2008}{\natexlab{a}}).

\bibitem[{\citenamefont{Voloshin and Zhang}(1996)}]{Voloshin:1994mz}
\bibinfo{author}{\bibfnamefont{S.}~\bibnamefont{Voloshin}} \bibnamefont{and}
  \bibinfo{author}{\bibfnamefont{Y.}~\bibnamefont{Zhang}}, \bibinfo{journal}{Z.
  Phys.} \textbf{\bibinfo{volume}{C70}}, \bibinfo{pages}{665}
  (\bibinfo{year}{1996}).

\bibitem[{\citenamefont{Kolb et~al.}(2001)\citenamefont{Kolb, Huovinen, Heinz,
  and Heiselberg}}]{Kolb:2000fha}
\bibinfo{author}{\bibfnamefont{P.~F.} \bibnamefont{Kolb}},
  \bibinfo{author}{\bibfnamefont{P.}~\bibnamefont{Huovinen}},
  \bibinfo{author}{\bibfnamefont{U.~W.} \bibnamefont{Heinz}}, \bibnamefont{and}
  \bibinfo{author}{\bibfnamefont{H.}~\bibnamefont{Heiselberg}},
  \bibinfo{journal}{Phys. Lett.} \textbf{\bibinfo{volume}{B500}},
  \bibinfo{pages}{232} (\bibinfo{year}{2001}).

\bibitem[{\citenamefont{Ollitrault}(1992)}]{Ollitrault:1992bk}
\bibinfo{author}{\bibfnamefont{J.-Y.} \bibnamefont{Ollitrault}},
  \bibinfo{journal}{Phys. Rev.} \textbf{\bibinfo{volume}{D46}},
  \bibinfo{pages}{229} (\bibinfo{year}{1992}).

\bibitem[{\citenamefont{Adcox et~al.}(2005)}]{Adcox:2004mh}
\bibinfo{author}{\bibfnamefont{K.}~\bibnamefont{Adcox}} \bibnamefont{et~al.}
  (\bibinfo{collaboration}{PHENIX}), \bibinfo{journal}{Nucl. Phys.}
  \textbf{\bibinfo{volume}{A757}}, \bibinfo{pages}{184} (\bibinfo{year}{2005}).

\bibitem[{\citenamefont{Back et~al.}(2005{\natexlab{c}})}]{Back:2004je}
\bibinfo{author}{\bibfnamefont{B.~B.} \bibnamefont{Back}} \bibnamefont{et~al.}
  (\bibinfo{collaboration}{PHOBOS}), \bibinfo{journal}{Nucl. Phys.}
  \textbf{\bibinfo{volume}{A757}}, \bibinfo{pages}{28}
  (\bibinfo{year}{2005}{\natexlab{c}}).

\bibitem[{\citenamefont{Adams et~al.}(2005{\natexlab{b}})}]{Adams:2005dq}
\bibinfo{author}{\bibfnamefont{J.}~\bibnamefont{Adams}} \bibnamefont{et~al.}
  (\bibinfo{collaboration}{STAR}), \bibinfo{journal}{Nucl. Phys.}
  \textbf{\bibinfo{volume}{A757}}, \bibinfo{pages}{102}
  (\bibinfo{year}{2005}{\natexlab{b}}).

\bibitem[{\citenamefont{Alver et~al.}(2010{\natexlab{a}})}]{Alver:2008gk}
\bibinfo{author}{\bibfnamefont{B.}~\bibnamefont{Alver}} \bibnamefont{et~al.}
  (\bibinfo{collaboration}{PHOBOS}), \bibinfo{journal}{Phys. Rev.}
  \textbf{\bibinfo{volume}{C81}}, \bibinfo{pages}{024904}
  (\bibinfo{year}{2010}{\natexlab{a}}).

\bibitem[{\citenamefont{Adams et~al.}(2006{\natexlab{b}})}]{Adams:2004pa}
\bibinfo{author}{\bibfnamefont{J.}~\bibnamefont{Adams}} \bibnamefont{et~al.}
  (\bibinfo{collaboration}{STAR}), \bibinfo{journal}{Phys. Rev.}
  \textbf{\bibinfo{volume}{C73}}, \bibinfo{pages}{064907}
  (\bibinfo{year}{2006}{\natexlab{b}}).

\bibitem[{\citenamefont{Ajitanand et~al.}(2005)}]{Ajitanand:2005jj}
\bibinfo{author}{\bibfnamefont{N.~N.} \bibnamefont{Ajitanand}}
  \bibnamefont{et~al.}, \bibinfo{journal}{Phys. Rev.}
  \textbf{\bibinfo{volume}{C72}}, \bibinfo{pages}{011902}
  (\bibinfo{year}{2005}).

\bibitem[{\citenamefont{Trainor and Kettler}(2008)}]{Trainor:2007fu}
\bibinfo{author}{\bibfnamefont{T.~A.} \bibnamefont{Trainor}} \bibnamefont{and}
  \bibinfo{author}{\bibfnamefont{D.~T.} \bibnamefont{Kettler}},
  \bibinfo{journal}{Int. J. Mod. Phys.} \textbf{\bibinfo{volume}{E17}},
  \bibinfo{pages}{1219} (\bibinfo{year}{2008}).

\bibitem[{\citenamefont{Adams et~al.}(2005{\natexlab{c}})}]{Adams:2005ph}
\bibinfo{author}{\bibfnamefont{J.}~\bibnamefont{Adams}} \bibnamefont{et~al.}
  (\bibinfo{collaboration}{STAR}), \bibinfo{journal}{Phys. Rev. Lett.}
  \textbf{\bibinfo{volume}{95}}, \bibinfo{pages}{152301}
  (\bibinfo{year}{2005}{\natexlab{c}}).

\bibitem[{\citenamefont{Adare et~al.}(2008{\natexlab{b}})}]{Adare:2007vu}
\bibinfo{author}{\bibfnamefont{A.}~\bibnamefont{Adare}} \bibnamefont{et~al.}
  (\bibinfo{collaboration}{PHENIX}), \bibinfo{journal}{Phys. Rev.}
  \textbf{\bibinfo{volume}{C77}}, \bibinfo{pages}{011901}
  (\bibinfo{year}{2008}{\natexlab{b}}).

\bibitem[{\citenamefont{Alver et~al.}(2010{\natexlab{b}})}]{Alver:2009id}
\bibinfo{author}{\bibfnamefont{B.}~\bibnamefont{Alver}} \bibnamefont{et~al.}
  (\bibinfo{collaboration}{PHOBOS}), \bibinfo{journal}{Phys. Rev. Lett.}
  \textbf{\bibinfo{volume}{104}}, \bibinfo{pages}{062301}
  (\bibinfo{year}{2010}{\natexlab{b}}).

\bibitem[{\citenamefont{Abelev et~al.}(2009{\natexlab{a}})}]{Abelev:2009qa}
\bibinfo{author}{\bibfnamefont{B.~I.} \bibnamefont{Abelev}}
  \bibnamefont{et~al.} (\bibinfo{collaboration}{STAR}), \bibinfo{journal}{Phys.
  Rev.} \textbf{\bibinfo{volume}{C80}}, \bibinfo{pages}{064912}
  (\bibinfo{year}{2009}{\natexlab{a}}), \eprint{0909.0191}.

\bibitem[{\citenamefont{Adamova et~al.}(2009)}]{Adamova:2009ah}
\bibinfo{author}{\bibfnamefont{D.}~\bibnamefont{Adamova}} \bibnamefont{et~al.}
  (\bibinfo{collaboration}{CERES}), \bibinfo{journal}{Phys. Lett.}
  \textbf{\bibinfo{volume}{B678}}, \bibinfo{pages}{259} (\bibinfo{year}{2009}).

\bibitem[{\citenamefont{Abelev et~al.}(2009{\natexlab{b}})}]{:2008nda}
\bibinfo{author}{\bibfnamefont{B.~I.} \bibnamefont{Abelev}}
  \bibnamefont{et~al.} (\bibinfo{collaboration}{STAR}), \bibinfo{journal}{Phys.
  Rev. Lett.} \textbf{\bibinfo{volume}{102}}, \bibinfo{pages}{052302}
  (\bibinfo{year}{2009}{\natexlab{b}}).

\bibitem[{\citenamefont{Wong}(2008)}]{Wong:2008yh}
\bibinfo{author}{\bibfnamefont{C.-Y.} \bibnamefont{Wong}},
  \bibinfo{journal}{Phys. Rev.} \textbf{\bibinfo{volume}{C78}},
  \bibinfo{pages}{064905} (\bibinfo{year}{2008}).

\bibitem[{\citenamefont{Pantuev}(2007)}]{Pantuev:2007sh}
\bibinfo{author}{\bibfnamefont{V.~S.} \bibnamefont{Pantuev}}
  (\bibinfo{year}{2007}), \eprint{arXiv:0710.1882}.

\bibitem[{\citenamefont{Gavin et~al.}(2009)\citenamefont{Gavin, McLerran, and
  Moschelli}}]{Gavin:2008ev}
\bibinfo{author}{\bibfnamefont{S.}~\bibnamefont{Gavin}},
  \bibinfo{author}{\bibfnamefont{L.}~\bibnamefont{McLerran}}, \bibnamefont{and}
  \bibinfo{author}{\bibfnamefont{G.}~\bibnamefont{Moschelli}},
  \bibinfo{journal}{Phys. Rev.} \textbf{\bibinfo{volume}{C79}},
  \bibinfo{pages}{051902} (\bibinfo{year}{2009}).

\bibitem[{\citenamefont{Dumitru et~al.}(2008)\citenamefont{Dumitru, Gelis,
  McLerran, and Venugopalan}}]{Dumitru:2008wn}
\bibinfo{author}{\bibfnamefont{A.}~\bibnamefont{Dumitru}},
  \bibinfo{author}{\bibfnamefont{F.}~\bibnamefont{Gelis}},
  \bibinfo{author}{\bibfnamefont{L.}~\bibnamefont{McLerran}}, \bibnamefont{and}
  \bibinfo{author}{\bibfnamefont{R.}~\bibnamefont{Venugopalan}},
  \bibinfo{journal}{Nucl. Phys.} \textbf{\bibinfo{volume}{A810}},
  \bibinfo{pages}{91} (\bibinfo{year}{2008}).

\bibitem[{\citenamefont{Ruppert and Renk}(2008)}]{Ruppert:2007mm}
\bibinfo{author}{\bibfnamefont{J.}~\bibnamefont{Ruppert}} \bibnamefont{and}
  \bibinfo{author}{\bibfnamefont{T.}~\bibnamefont{Renk}},
  \bibinfo{journal}{Acta Phys. Polon. Supp.} \textbf{\bibinfo{volume}{1}},
  \bibinfo{pages}{633} (\bibinfo{year}{2008}).

\bibitem[{\citenamefont{Pruneau et~al.}(2008)\citenamefont{Pruneau, Gavin, and
  Voloshin}}]{Pruneau:2007ua}
\bibinfo{author}{\bibfnamefont{C.~A.} \bibnamefont{Pruneau}},
  \bibinfo{author}{\bibfnamefont{S.}~\bibnamefont{Gavin}}, \bibnamefont{and}
  \bibinfo{author}{\bibfnamefont{S.~A.} \bibnamefont{Voloshin}},
  \bibinfo{journal}{Nucl. Phys.} \textbf{\bibinfo{volume}{A802}},
  \bibinfo{pages}{107} (\bibinfo{year}{2008}).

\bibitem[{\citenamefont{Hwa}(2009)}]{Hwa:2009bh}
\bibinfo{author}{\bibfnamefont{R.~C.} \bibnamefont{Hwa}}
  (\bibinfo{year}{2009}), \eprint{arXiv:0904.2159}.

\bibitem[{\citenamefont{Takahashi et~al.}(2009)}]{Takahashi:2009na}
\bibinfo{author}{\bibfnamefont{J.}~\bibnamefont{Takahashi}}
  \bibnamefont{et~al.}, \bibinfo{journal}{Phys. Rev. Lett.}
  \textbf{\bibinfo{volume}{103}}, \bibinfo{pages}{242301}
  (\bibinfo{year}{2009}).

\bibitem[{\citenamefont{Nagle}(2009)}]{Nagle:2009wr}
\bibinfo{author}{\bibfnamefont{J.~L.} \bibnamefont{Nagle}},
  \bibinfo{journal}{Nucl. Phys.} \textbf{\bibinfo{volume}{A830}},
  \bibinfo{pages}{147c} (\bibinfo{year}{2009}).

\bibitem[{\citenamefont{Sorensen}(2010)}]{Sorensen}
\bibinfo{author}{\bibfnamefont{P.}~\bibnamefont{Sorensen}},
  \bibinfo{journal}{to appear in J. Phys.} \textbf{\bibinfo{volume}{G}}
  (\bibinfo{year}{2010}), \eprint{arXiv:1002.4878v1}.

\bibitem[{\citenamefont{Mishra et~al.}(2008)\citenamefont{Mishra, Mohapatra,
  Saumia, and Srivastava}}]{Mishra:2007tw}
\bibinfo{author}{\bibfnamefont{A.~P.} \bibnamefont{Mishra}},
  \bibinfo{author}{\bibfnamefont{R.~K.} \bibnamefont{Mohapatra}},
  \bibinfo{author}{\bibfnamefont{P.~S.} \bibnamefont{Saumia}},
  \bibnamefont{and} \bibinfo{author}{\bibfnamefont{A.~M.}
  \bibnamefont{Srivastava}}, \bibinfo{journal}{Phys. Rev.}
  \textbf{\bibinfo{volume}{C77}}, \bibinfo{pages}{064902}
  (\bibinfo{year}{2008}).

\bibitem[{\citenamefont{Lin et~al.}(2005)\citenamefont{Lin, Ko, Li, Zhang, and
  Pal}}]{Lin:2004en}
\bibinfo{author}{\bibfnamefont{Z.-W.} \bibnamefont{Lin}},
  \bibinfo{author}{\bibfnamefont{C.~M.} \bibnamefont{Ko}},
  \bibinfo{author}{\bibfnamefont{B.-A.} \bibnamefont{Li}},
  \bibinfo{author}{\bibfnamefont{B.}~\bibnamefont{Zhang}}, \bibnamefont{and}
  \bibinfo{author}{\bibfnamefont{S.}~\bibnamefont{Pal}},
  \bibinfo{journal}{Phys. Rev.} \textbf{\bibinfo{volume}{C72}},
  \bibinfo{pages}{064901} (\bibinfo{year}{2005}).

\bibitem[{\citenamefont{Abelev et~al.}(2008)}]{Abelev:2008un}
\bibinfo{author}{\bibfnamefont{B.~I.} \bibnamefont{Abelev}}
  \bibnamefont{et~al.} (\bibinfo{collaboration}{STAR}),
  \bibinfo{journal}{submitted to Phys. Rev.} \textbf{\bibinfo{volume}{C}}
  (\bibinfo{year}{2008}), \eprint{0806.0513}.

\bibitem[{\citenamefont{PHOBOS}({\natexlab{a}})}]{Phobosdata1}
\bibinfo{author}{\bibnamefont{PHOBOS}},
  \urlprefix\url{http://www.phobos.bnl.gov/Publications/Physics/Trig_Correl/in%
dex.htm}.

\bibitem[{\citenamefont{PHOBOS}({\natexlab{b}})}]{Phobosdata2}
\bibinfo{author}{\bibnamefont{PHOBOS}},
  \urlprefix\url{http://www.phobos.bnl.gov/Publications/Physics/AA_2part_corre%
l/index.htm}.

\bibitem[{\citenamefont{STAR}()}]{Stardata1}
\bibinfo{author}{\bibnamefont{STAR}}, \emph{\bibinfo{title}{private
  communication}},
  \urlprefix\url{http://drupal.star.bnl.gov/STAR/publications/charge-independe%
nt-ci-and-charge-dependent-cd\ -correlations-function-centrality-formed-delta\
  -phi-delta-eta-charged-p}.

\bibitem[{\citenamefont{Alver et~al.}(2008{\natexlab{a}})}]{Alver:2008zza}
\bibinfo{author}{\bibfnamefont{B.}~\bibnamefont{Alver}} \bibnamefont{et~al.}
  (\bibinfo{collaboration}{PHOBOS}), \bibinfo{journal}{Phys. Rev.}
  \textbf{\bibinfo{volume}{C77}}, \bibinfo{pages}{014906}
  (\bibinfo{year}{2008}{\natexlab{a}}).

\bibitem[{\citenamefont{Alver et~al.}(2010{\natexlab{c}})}]{Alver:2010rt}
\bibinfo{author}{\bibfnamefont{B.}~\bibnamefont{Alver}} \bibnamefont{et~al.}
  (\bibinfo{collaboration}{PHOBOS}), \bibinfo{journal}{Phys. Rev.}
  \textbf{\bibinfo{volume}{C81}}, \bibinfo{pages}{034915}
  (\bibinfo{year}{2010}{\natexlab{c}}).

\bibitem[{\citenamefont{Alver et~al.}(2008{\natexlab{b}})\citenamefont{Alver,
  Baker, Loizides, and Steinberg}}]{Alver:2008aq}
\bibinfo{author}{\bibfnamefont{B.}~\bibnamefont{Alver}},
  \bibinfo{author}{\bibfnamefont{M.}~\bibnamefont{Baker}},
  \bibinfo{author}{\bibfnamefont{C.}~\bibnamefont{Loizides}}, \bibnamefont{and}
  \bibinfo{author}{\bibfnamefont{P.}~\bibnamefont{Steinberg}}
  (\bibinfo{year}{2008}{\natexlab{b}}), \eprint{arXiv:0805.4411}.

\bibitem[{\citenamefont{Ma et~al.}(2006)}]{Ma:2006fm}
\bibinfo{author}{\bibfnamefont{G.~L.} \bibnamefont{Ma}} \bibnamefont{et~al.},
  \bibinfo{journal}{Phys. Lett.} \textbf{\bibinfo{volume}{B641}},
  \bibinfo{pages}{362} (\bibinfo{year}{2006}).

\bibitem[{\citenamefont{Ma et~al.}(2008)}]{Ma:2008nd}
\bibinfo{author}{\bibfnamefont{G.~L.} \bibnamefont{Ma}} \bibnamefont{et~al.},
  \bibinfo{journal}{Eur. Phys. J.} \textbf{\bibinfo{volume}{C57}},
  \bibinfo{pages}{589} (\bibinfo{year}{2008}).

\bibitem[{\citenamefont{Zhang et~al.}(2007)}]{Zhang:2007qx}
\bibinfo{author}{\bibfnamefont{S.}~\bibnamefont{Zhang}} \bibnamefont{et~al.},
  \bibinfo{journal}{Phys. Rev.} \textbf{\bibinfo{volume}{C76}},
  \bibinfo{pages}{014904} (\bibinfo{year}{2007}).

\bibitem[{\citenamefont{Li et~al.}(2009)}]{Li:2009ti}
\bibinfo{author}{\bibfnamefont{W.}~\bibnamefont{Li}} \bibnamefont{et~al.},
  \bibinfo{journal}{Phys. Rev.} \textbf{\bibinfo{volume}{C80}},
  \bibinfo{pages}{064913} (\bibinfo{year}{2009}).

\bibitem[{\citenamefont{Gyulassy and Wang}(1994)}]{Gyulassy:1994ew}
\bibinfo{author}{\bibfnamefont{M.}~\bibnamefont{Gyulassy}} \bibnamefont{and}
  \bibinfo{author}{\bibfnamefont{X.-N.} \bibnamefont{Wang}},
  \bibinfo{journal}{Comput. Phys. Commun.} \textbf{\bibinfo{volume}{83}},
  \bibinfo{pages}{307} (\bibinfo{year}{1994}).

\bibitem[{\citenamefont{Abelev et~al.}(2009{\natexlab{c}})}]{Abelev:2009jv}
\bibinfo{author}{\bibfnamefont{B.~I.} \bibnamefont{Abelev}}
  \bibnamefont{et~al.} (\bibinfo{collaboration}{STAR})
  (\bibinfo{year}{2009}{\natexlab{c}}), \eprint{arXiv:0912.3977}.

\end{thebibliography}

\end{document}